\documentclass{aa} 

\pdfoutput=1
\usepackage{graphicx}
\usepackage{txfonts} \usepackage{color} \usepackage{enumitem}
\usepackage{lineno}

\usepackage[%
  breaklinks = true,
  colorlinks = true,
  urlcolor   = blue,
  citecolor  = blue,
  linkcolor  = blue,
]{hyperref}

\usepackage{bm}
\usepackage{ulem}

\begin{document} 



\def\dps{\displaystyle}
\def\scs{\scriptstyle}
\def\scscs{\scriptscriptstyle}

\newcommand{\pderdps}[2]{\fracdps{\partial\hskip 1pt \strut
    #1}{\partial\hskip 1pt \strut #2}}
\newcommand{\tderdps}[2]{{\fracdps{d\hskip 0.pt \strut #1}{d\hskip 1pt \strut #2}}}

\newcommand{\fracdps}[2]{{{\strut\dps#1}\over{\strut\dps #2}}}

\def\difvar{B}

\newcommand{\dotBzero}{{\dot{\difvar}_0}}
\newcommand{\Bzero}{{\difvar_0}}
\newcommand{\Bzerosq}{{\difvar_0^2}}

\newcommand{\Psivar}{\Theta}
\newcommand{\Ktilde}{{\hat{\it K}}}
\newcommand{\Khat}{\Ktilde}
\newcommand{\xitilde}{{\hat{\xi}}}
\newcommand{\Rfront}{{L}}
\newcommand{\Rfrontsq}{{L^2}}
\newcommand{\Rfrontdot}{{\dot{L}}}

\newcommand{\Rcs}{{L_{cs}^{}}}

\newcommand{\xmax}{{x_{max}^{}}^{}}
\newcommand{\ymax}{{y_{max}^{}}^{}}
\newcommand{\xhat}{{\hat{x}}}
\newcommand{\pot}[2]{{#1\,10^{#2}}}

\newcommand{\tzz}{t_{00}}
\newcommand{\taudzz}{{\tau_d}_{00}^{}}
\newcommand{\Bzz}{B_{00}}
\newcommand{\Bzzsq}{B^2_{00}}
\newcommand{\Rfrontzz}{\Rfront_{00}}
\newcommand{\Rfrontzzsq}{\Rfront^2_{00}}
\newcommand{\xitildejunction}{\xitilde_r}

\def\defdef{{\buildrel \rm def \over =}}

\renewcommand{\vec}[1]{\mathbf{#1}}

\newcommand{\uvec}[1]{{\vec{e}_#1^{}}}
\newcommand{\src}{\ssign_{{\Rfront^{}}}}
\newcommand{\xvec}{{\vec{x}}}
\newcommand{\Bvec}{{\vec{B}}}
\newcommand{\Evec}{{\vec{E}}}
\newcommand{\jvec}{{\vec{j}}}
\newcommand{\curl}{{\hbox{\hskip 2truept\bf curl\hskip 1.5truept}}}
\newcommand{\uuvec}{{\vec{u}}}
\newcommand{\sigmaohm}{{\sigma_{el}^{}}}
\newcommand{\muzero}{{\mu_0^{}}}
\newcommand{\chiamb}{{\chi_{a}}}
 \newcommand{\etaamb}{{\eta_{a}^{}}}

\newcommand{\diffflux}{{F_{d}}}
\newcommand{\difffluxt}{{\tilde{F}_{d}}}
\newcommand{\difffluxnodim}{{D}}


\title{Ambipolar diffusion: \\
Self-similar solutions and MHD code testing.}
\subtitle{Cylindrical symmetry}

\author{F.~Moreno-Insertis\inst{1,2}
\and
D.~N\'obrega-Siverio\inst{1,2,3,4}
\and
E.~R.~Priest\inst{5}
\and
A.~W.~Hood\inst{5}
}

\institute{Instituto de Astrofisica de Canarias, 38205 La Laguna, Tenerife,
  Spain\\ \email{fmi@iac.es}\\ \and Departamento de Astrofisica, Universidad
  de La Laguna, 38205 La Laguna, Tenerife, Spain. \\ \and Rosseland Centre
  for Solar Physics, University of Oslo, PO Box 1029 Blindern, NO-0315 Oslo,
  Norway \\ \and Institute of Theoretical Astrophysics, University of Oslo,
  PO Box 1029 Blindern, 0315 Oslo, Norway\\ \and School of Mathematics and
  Statistics, University of St Andrews, St Andrews KY16 9SS, UK\\}

\vspace{1cm}

\abstract
{Ambipolar diffusion is a process occurring in partially ionised
  astrophysical systems that imparts a complicated mathematical and physical
  nature to Ohm's law. The numerical codes that solve the
  magnetohydrodynamic (MHD) equations have to be able to deal with the
  singularities that are naturally created in the system by the ambipolar
  diffusion term.}
{The global aim is to calculate a set of theoretical self-similar
    solutions to the nonlinear diffusion equation with cylindrical symmetry
    that can be used as tests for MHD codes which include the
    ambipolar diffusion term.}
{First, following the general methods developed in the applied mathematics
  literature, we obtained the theoretical solutions as eigenfunctions of a
  nonlinear ordinary differential equation.  Phase-plane techniques were used
  to integrate through the singularities at the locations of the nulls, which
  correspond to infinitely sharp current sheets.  In the second half of the
  paper, we consider the use of these solutions as tests for MHD codes. To
  that end, we used the Bifrost code, thereby testing the capabilities of
  these solutions as tests as well as (inversely) the accuracy of Bifrost's
  recently developed ambipolar diffusion module.  }
{The obtained solutions are shown to constitute a demanding, but nonetheless
  viable, test for MHD codes that incorporate ambipolar diffusion. Detailed
  tabulated runs of the solutions have been made available at a public
  repository. The 
  Bifrost code is able to reproduce the theoretical solutions with sufficient
  accuracy up to very advanced diffusive times. Using the code, we also
  explored the asymptotic properties of our theoretical solutions in time when
  initially perturbed with either small or finite perturbations.}
{The functions obtained in this paper are relevant as physical solutions and also as tests for general MHD codes. They provide a more stringent and general test than the simple Zeldovich-Kompaneets-Barenblatt-Pattle solution.  }
\keywords{Magnetic fields -- Plasmas -- Methods: numerical -- Methods: analytical -- Diffusion }
\maketitle

\section{Introduction \label{sec:intro}}

Ambipolar diffusion is an important, and at times crucial process at work in magnetised plasmas in which ionised and neutral
populations coexist. For a long time, and increasingly in recent years,
numerical models of the time-evolution of partially ionised plasmas that
include ambipolar diffusion have been constructed. However, the ambipolar diffusion
  coefficient is proportional to $B^2$, and so, by its own nature, it
can lead to the formation of sharp, mathematically singular current
sheets in the neighbourhood of magnetic null points \citep{Zweibel_1994,
  Brandenburg_Zweibel_1994}. From the point of view of applied mathematics,
singularities are a natural outcome of the evolution of nonlinear
  diffusion equations (see the book by \citealt{Vazquez_2007}). 
Interestingly, as detailed in the present paper, in the
  ambipolar diffusion case, the passage of magnetic flux across the singular
current sheet occurs at a finite rate through the combination of the
  vanishing diffusion coefficient and the infinite slope of the magnetic
  profile.  Now, if the time evolution of a system leads to the spontaneous
formation of essential singularities that play a relevant role in the
system, then the numerical solution of the problem may become quite
complicated.  Thus, a question arises with regard to the capabilities of
state-of-the-art MHD numerical codes in coping with the expected singular
current sheets and, in addition, whether it is possible to provide a suite of tests to probe
their proficiency when dealing with such singularities.

The plasma in the low atmosphere of the Sun and other cool stars, in the
interstellar medium, accretion disks, planetary ionospheres, and in various
other cosmic environments is partially ionised -- in fact, it is sometimes
only weakly ionised \citep[see, e.g. the reviews from][]{Shu_etal_1987,
  McKee_Ostriker_ARAA_2007,Zweibel_etal_2011, Crutcher_2012, Leake_etal_2014,
  Zweibel_2015, Ballester_2018}. For example, in crucial layers of the low
solar atmosphere (e.g. the high photosphere and chromosphere), the standard
VALC model \citep{val1981} predicts the ionisation degree of Hydrogen to be
below $10^{-2}$. In dynamical models, such as those calculated with the
radiation-MHD Bifrost code \citep[e.g.][]{Carlsson_etal_2016}, the horizontal
averages of the ratio between electron and Hydrogen number density, $\langle
n_{el}/n_H \rangle$, are also typically below $10^{-2}$ at those heights, in
spite of the juxtaposition of cool pockets and hotter structures apparent in
the 2D and 3D realistic numerical models that include those layers
\citep[see, e.g.][]{Wedemeyer_etal_2004, Hansteen_etal_procs_2007,
  Leenaarts_etal_2007, Leenaarts_etal_2011, Nobrega-Siverio_etal_2020a}.  In
a magnetised, partially ionised plasma, it is only the charged species that
are subject to the Lorentz force: this causes a drift between the neutrals
and the electron-ion gas, which is countered by their mutual friction due to
collisions, or by charge-exchange phenomena, or by a combination of
both. When the neutral-charged coupling is strong enough, however, the plasma
can still be treated as a single fluid, provided the standard Ohm's law used
in simple magnetohydrodynamics is augmented with extra terms
\citep{cowling1957magnetohydrodynamics, braginskii_1965, Mitchner:1973}.

Following those basic references, in the simplest case such a
  generalised Ohm's law can be written in the following form:
\begin{equation}\label{eq:ohm_ambipolar}
\Evec\; +\; \uuvec\times \Bvec \;= \fracdps{\jvec}{\sigmaohm} \;+\;
\fracdps{\jvec\times\Bvec}{e\,n_e}\; +\;
\chiamb\,B^2\; \muzero\,\jvec_\perp \;,
\end{equation}

\noindent 
where $\uuvec$ is the centre-of-mass velocity of the plasma,
  $\jvec_\perp$ is the component of $\jvec$ perpendicular to $\Bvec$,
  $\sigmaohm$ is the electrical conductivity, $e$ and $n_e$ are the electric
  charge and number density of the electrons, respectively, and $\chiamb$ is
a material coefficient, namely:
\begin{eqnarray}
\chiamb &\defdef& \fracdps{1}{\muzero} \;
\left(\fracdps{\rho_n}{\rho}\right)^2\,\fracdps {1}{
         \,n_e\,m_{in}\,\nu_{in}}\;\;, \hfill
\end{eqnarray}
with $\rho$ and $\rho_n$ the total and neutral mass density, respectively, $\nu_{in}$ the ion-neutral
collision frequency and $m_{in}$ the reduced ion-neutral mass.  The last two
terms in Eq.~(\ref{eq:ohm_ambipolar}) are the Hall term and the
ambipolar term, respectively. To focus on the ambipolar diffusion
problem, we can consider the case when the bulk plasma speed is zero (e.g.
because the neutrals are at rest and the ionisation level is low) and the
Hall and electric conductivity terms are negligible. In that case, one is
left with the left-most and right-most terms in
(\ref{eq:ohm_ambipolar}) alone.  Using Faraday's induction law, we obtain
the corresponding equation for the evolution of the magnetic field:
\begin{equation}\label{eq:induc_ambipolar}
\pderdps{\Bvec}{t} = -\curl \left[\; \etaamb\, (\curl\Bvec)_\perp\, \right]
\;,
\end{equation}
with $\etaamb$, the ambipolar diffusion coefficient, defined by
\begin{equation}\label{eq:chiamb}
\etaamb \;\defdef \; \chiamb\,B^2\,.
\end{equation}
Equations (\ref{eq:induc_ambipolar}) and (\ref{eq:chiamb}) constitute the
pure ambipolar diffusion problem. The term "diffusion" in this context is
standard in the literature, even though it is not clear that a general
magnetic field configuration governed by
Eqs.~(\ref{eq:induc_ambipolar}) -- (\ref{eq:chiamb}) should show what would
intuitively be associated with simple diffusive behaviour.  On the one hand,
$\etaamb$ depends on $B^2$; thus, even the simplest diffusive behaviour we can
expect from Eqs.~(\ref{eq:induc_ambipolar}) -- (\ref{eq:chiamb}) (e.g. when the
field is parallel to itself everywhere) is intrinsically nonlinear.  Furthermore,
in a general case, the magnetic field has two
or three nonzero components and $\jvec_\perp$ is different from $\jvec$, so
Eq.~(\ref{eq:induc_ambipolar}) inextricably entangles the field
components in a vector equation that is not a simple diffusion equation.
There is extensive literature on this topic, starting with the seminal paper by
\citet{Mestel_Spitzer_1956} and encompassing hundreds of papers, that studies
various aspects of the ambipolar diffusion problem from a physical and
astrophysical point of view; for more details, we refer to the reviews mentioned
at the beginning of this section.  Of particular interest with regard to the present
article are those papers studying the sharp electric current sheets that form
when the magnetic field profile approaches a null point, such as those
  of \citet{Zweibel_1994} and \citet{Brandenburg_Zweibel_1994} mentioned
  above. These authors have shown how the 1D $B$
profiles governed by the ambipolar diffusion equation and containing a null
tend to develop a singular shape in time of the form $|\hskip 1pt B\hskip
1pt| \propto |\hskip 1pt \delta \hskip 1pt|^{1/3}$, with $\delta$ the
distance to the null, which could then remain as a stationary solution. They
also showed how a differentially rotating magnetic slab tends to develop
sharp current sheets between rotating layers -- again because of the effect of
ambipolar diffusion. Later papers showed how such singular profiles may
affect the process of magnetic cancellation and reconnection in them
\citep{Brandenburg_Zweibel_1995, Heitsch_Zweibel_2003a,
  Heitsch_Zweibel_2003b}.

In this paper we concentrate on a simple situation in which
Eq.~(\ref{eq:induc_ambipolar}) leads to a nonlinear diffusion equation,
namely, 
the case when $\Bvec$ points everywhere in the same direction, such as, for
instance, 
along the $z$-axis, $\Bvec = B(x,y, t)\, \uvec{z}$, for which the equation
becomes
\begin{equation}\label{eq:diffus_scalar}
\pderdps{B}{t} = \nabla \cdot \left(\,\chiamb\, B^2\,\nabla B\right)\;.
\end{equation}
We aim to determine explicit self-similar solutions for this equation when
$\chiamb$ is a constant; given the constraints, there are relevant cases in
one and two dimensions, namely, when the system has either axial symmetry (the
cylindrical case) or mirror symmetry (the plane-parallel
  or Cartesian case). Here, we are dealing with the
  cylindrical case, while the plane-parallel case will be treated in a follow-up
  study. Using coordinates $(r, \varphi, z)$, the equation we solve
  here is therefore: 
\begin{equation}\label{eq:diffusion}
  \pderdps{\difvar}{t} \;=
  \;\fracdps{\chi_a}{r}\,\pderdps{}{r}\left(r\,\difvar^2\,
  \pderdps{\difvar}{r} \right)\,.
\end{equation}
Equations (\ref{eq:diffus_scalar}) and (\ref{eq:diffusion}) 
are of diffusion type with nonlinear diffusion
coefficient $\chi_a\,\difvar^2$. 

Diffusion problems in which the diffusion coefficient is proportional to a
positive power (e.g.,~2,~or~5/2) of the diffusing quantity are important in
many different fields of physics in addition to ambipolar diffusion, such as
heat conduction in hot plasmas, flows in porous media, boundary layers, thin
liquid film spreading, and many others (see the book by
\citealt{Vazquez_2007}). In early studies \citep{Zeldovich_Kompaneets_1950,
  Barenblatt1952, Pattle1959}, fundamental self-similar solutions were found
in Cartesian, cylindrical, and spherical coordinates: we collectively
refer to them as the Zeldovich-Kompaneets-Barenblatt-Pattle solutions (or ZKBP, for
short\footnote{These solutions are variously known in the mathematical and
  physical literature as the Barenblatt solution, the Zeldovich-Kompaneets
  solution, the Barenblatt-Pattle solution or the Zeldovich-Barenblatt-Pattle
  solution. In this paper, we simply combine those designations into the
  acronym ZKBP, thus preserving the historical order}).  These solutions have
compact support (i.e. the set of points where they are nonzero is
  contained within a finite distance of the origin) and are unsigned,
that is, they do not have any internal nulls; the outer boundary has infinite
slope and expands in time with a speed which is a natural combination of the
parameters of the problem. Since those early works, the properties of the
self-similar solutions of the nonlinear diffusion equation have been studied
with great generality using phase-plane techniques for the spatial part (see,
e.g. \citealt{Grundy1979}, \citealt{Hulshof_1991} and the encompassing
review by \citealt{Vazquez_2007}). In such works, similarity solutions were
considered for the so-called 'porous medium equation', both for the
1D and for the radially-symmetric multidimensional cases. A general way of writing that equation is the following:

\begin{equation}\label{eq:porous_medium_equation}
\pderdps{u}{t} = \fracdps{1}{m} \Delta\left(|u|^{m-1}\,u\right)\;,
\end{equation}
with $u(x, t)$ a scalar function of the radial coordinate $x=|\hskip 1pt
\xvec\hskip 1pt|$; $\xvec$ the radial vector in $n\ge 1$ dimensions; 
$m$ a fixed parameter generally fulfilling $m \ge 1$;
and $\Delta$ the $n$-dimensional Laplacian operator.  Except for a constant,
the particular case for $m=3$ and  two dimensions coincides with
Eq.~(\ref{eq:diffusion}). The results of
\citet{Hulshof_1991} are of particular interest: that author considered
solutions with sign changes, that is, 
internal nulls, within their domain, but otherwise of the standard
self-similar form $u(x,t) = t^\alpha U(\eta)$, with $\eta = x\, t^{-\gamma}$;
$\alpha$ and $\gamma$ two constants; and $U(\eta)$ the (undetermined) spatial
part for the solution. For the case of a solution that expands in space and
decays in time ($\alpha<0$ and $\gamma>0$), he proved the existence of a
countable series of self-similar solutions with compact support, all with a
finite number of internal nulls; the first solution in the series is the ZKBP
solution, which has no nulls, and all the others have one internal null more
than the preceding one in the series. The exponents $\alpha$ and $\gamma$ in
the series are ordered such that the solutions decay faster but expand more
slowly the greater the number of nulls in their interior. In his paper,
\citet{Hulshof_1991} provides this classification, proves the mathematical
theorems supporting it, and studies the phase-plane properties of the
solutions; however, their specific spatial shapes $U(\eta)$ are not shown.

In this paper, we are interested not only in the mathematical
  properties of the ambipolar diffusion as a nonlinear diffusion process,
  but also in the inclusion of ambipolar diffusion terms in MHD codes.
  In astrophysics, over the past few decades, multidimensional MHD
  computer codes have been developed that model a variety of physical
  processes including ambipolar diffusion. Representative examples of such
  codes and simulations outside solar physics can be found in
  \citet{Basu_Mouschovias_1994, MacLow:1995, Padoan_etal_2000,
    Basu_Ciolek_2004, Kudoh_Basu_2008, Choi_etal_2009, Gressel_etal_2015,
    Tomida_etal_2015, osullivan2007, Masson_etal_2012, Vigano2019}; and
    \citet{Grassi:2019}.  The consideration of ambipolar diffusion processes in
  solar physics started many decades ago \citep[e.g.][]{Parker63}, but it has
  undergone a true explosion in terms of its use in large numerical models
  \citep[e.g.][]{Leake:2005rt,leake06,Arber_etal_2007,Cheung_Cameron_2012,
    leake_linton_2013, martinez_sykora_etal_partial_ionization_2012,
    Martinez-Sykora_etal_2017a, Martinez-Sykora_etal_2017_science,
    Martinez-Sykora_etal_2020_neq, Martinez-Sykora_etal_2020, Ni_etal_2015,
    Ni_etal_2016, Ni_etal_2021, Khomenko_etal_2017,
    Khomenko_etal_2018,Khomenko_etal_2021, Gonzalez-Morales_2018,
    Gonzalez-Morales_2020, Nobrega-Siverio_etal_2020b,
    Nobrega-Siverio_etal_2020a, Popescu_Keppens_2021}.
Such numerical calculations often encounter a problem: given the
comparatively high values of $\chiamb$ in different cosmic environments, the
advance in time may grind to a halt in magnetised regions when a standard
Courant-Friedrichs-Lewy condition is adopted for the timestep based on
$\etaamb$. The recent papers by \citet{Gonzalez-Morales_2018} and
\citet{Nobrega-Siverio_etal_2020b} describe the construction of so-called
super-time-stepping (STS) modules for the Mancha code and for the Bifrost
code, respectively, designed with the aim to overcome that stiffness problem. In at
least three of the papers cited above \citep{Masson_etal_2012, Vigano2019,
  Nobrega-Siverio_etal_2020b}, the basic ZKBP solution in cylindrical
coordinates was used to test the ambipolar diffusion module, given its
simplicity and the analytical expression available for it. However, as
already mentioned, the ambipolar diffusion problem tends to give rise to
sharp current sheets and singularities. The ZKBP solution is comparatively
smooth in that sense and it would be good to have some other
canonical solutions on hand that include current sheets having higher degrees of singularity,
which naturally occur in the ambipolar diffusion problem.

The objective of this paper is to calculate signed self-similar solutions of
Eq.~(\ref{eq:diffusion}) with compact support and propose that they be used as simple, but
nonetheless demanding tests for the ambipolar diffusion modules in MHD codes that
include the generalised Ohm's law.  First, we obtain, through separation of
variables and numerical calculation, a set of such solutions.  As expected
from the general considerations above, they have a finite number of nulls
within their domain which turn out to be locations of unavoidable
singularities; hence, we devised a method that permits a smooth calculation
of the solutions across the singularities by numerical means. Secondly, we used
some of those solutions to test the capabilities of the new ambipolar
diffusion module for the Bifrost code implemented by
\citet{Nobrega-Siverio_etal_2020b}, thereby showing that these solutions
constitute a suitable test for benchmarking MHD codes.

The layout of the paper is: in Sect.~\ref{sec:self-similar}, we describe the basic equation, the boundary conditions, and the method to calculate the solutions
numerically for the cylindrical case. The corresponding
solutions are presented and briefly discussed in
Sect.~\ref{sec:solutions_section}.
Then, in Sect.~\ref{sec:tests}, the use of all those solutions as tests for
numerical codes is explained on the basis of the specific example in the
Bifrost code. Sect.~\ref{sec:discussion} presents a discussion and 
conclusions.

\section{Finding self-similar solutions}\label{sec:self-similar}

\subsection{The self-similar ansatz}\label{sec:self-similar_ansatz}

To find self-similar solutions of the nonlinear diffusion equation for the
case with cylindrical symmetry, Eq.~(\ref{eq:diffusion}), we first note that,
just 
as in the case of heat flux in the nonlinear heat conduction equation, we
can here speak of a diffusive flux given by
\begin{equation}\label{eq:diffusive_flux}
\diffflux \,\defdef - \chi_a\,\difvar^2\,\pderdps{\difvar}{r}\;,
\end{equation}
which tends to carry magnetic field from more strongly magnetised regions to
weaker-field ones.
Now, for the separation of variables, we follow standard applied mathematical
procedures \citep[e.g.][]{Vazquez_2007}, by seeking solutions of the form
\begin{equation}\label{eq:self-similar_ansatz}
  \difvar(r,t) = \Bzero(t) \; f(\xi)\;, \quad\hbox{with}\quad \xi(r,t) =
  \fracdps{r}{\Rfront(t)}\;,
\end{equation}
and with $\Rfront(t)$ and $\Bzero(t)$ being global scales for the spatial
  coordinate and for the magnetic field, respectively.  For simplicity, we
  choose $f(\xi=0) = 1$, so $\Bzero(t)$ corresponds to the
  value of the solution at the cylindrical axis.  If a dot and a prime
indicate total derivatives with respect to $t$ and $\xi$, respectively,
Eq.~(\ref{eq:diffusion}) becomes:
\begin{equation}\label{eq:after_selfsimilar_2}
0 = - \left(\fracdps{\Rfront}{\Bzero}\right)^2 \fracdps{\dotBzero}{\Bzero}\,f
\,+\, \left(\fracdps{\Rfront}{\Bzero}\right)^2 \fracdps{\Rfrontdot}{\Rfront}
\,\xi \,f^\prime \,+\, \fracdps{\chi_a}{\xi}\,\left(\strut
\xi\,f^2\,f^\prime\right)^\prime\;,
\end{equation}
while the diffusive flux defined by Eq.~(\ref{eq:diffusive_flux}) takes
the form: 
\begin{equation}\label{eq:diffusive_flux_separation}
\diffflux = \chi_a\, \fracdps{\Bzero^3}{\Rfront}\; \difffluxnodim \;,
\quad\hbox{with}\quad \difffluxnodim\,\defdef - f^2 f'.
\end{equation}
To proceed, all variables are normalised by choosing values for $\Bzero$ and
$\Rfront$ at a given initial time, $\tzz$, say $\Bzz$ and $\Rfrontzz$. Also,  a natural time unit is defined to be $\Rfrontzzsq/(\chi_a\,\Bzzsq)$, which is equivalent
to setting $\chi_a=1$ in the foregoing equations.  Then we request that all
terms in Eq.~(\ref{eq:after_selfsimilar_2}) be time-independent by setting
\begin{equation}\label{eq:K_def}
- \left(\fracdps{\Rfront}{\Bzero}\right)^2 \fracdps{\dotBzero}{\Bzero} \;=\;
K\;, \qquad \left(\fracdps{\Rfront}{\Bzero}\right)^2
\fracdps{\Rfrontdot}{\Rfront} \;=\;H\;,
\end{equation}
with $K$ and $H$ two constants which are positive for decreasing $\Bzero$ and
growing $\Rfront$, respectively.  From Eq.~(\ref{eq:K_def}), we have:
\begin{equation}\label{eq:B_R_constraint}
\Bzero(t) \; =\; \left[\tau_d(t)\right]^{\hskip 2pt \alpha}\;,\quad
\Rfront(t) \; =\; \left[\tau_d(t)\right]^{\hskip 2pt \gamma}\;,
\end{equation}
with
\begin{equation}\label{eq:exponents}
\alpha \;\defdef\; \frac{-K}{ 2 K + 2 H} \;, \quad
\gamma \;\defdef\; \frac{H}{2 K + 2 H} \;,
\end{equation}
and $\tau_d(t)$ is an evolving timescale for the diffusion problem:
\begin{equation}\label{eq:tau_d}
\tau_d(t) \;\defdef\; 1 \,+\, 2 \,(H+K)\, (t-\tzz)\;.
\end{equation}
Using $K$ and $H$, Eq.~(\ref{eq:after_selfsimilar_2}) becomes an equation
for the spatial part of $B$, namely,
\begin{equation}\label{eq:selfsimilar_0}
  0\;=\; K\,f \,+\, H\,\xi\,f^\prime \,+\, \fracdps{1}{\xi}\left(
  \xi\,f^2\,f^\prime\right)^\prime \;.
\end{equation}
To find self-similar solutions we must therefore solve
Eq.~(\ref{eq:selfsimilar_0}) subject to appropriate boundary conditions. This
equation is akin to those found for the general problem in the mathematical
literature mentioned in the introduction.
In fact, introducing
$\tau_d$ as time variable instead of $t$ would bring our solutions to the
standard self-similar form 
$B(r,\tau) = \tau^{\hskip  2pt \alpha}\; f\left(r\,\tau^{\hskip 1pt
  -\gamma}\right)$ \citep[e.g.][]{Zeldovich_book1967, Vazquez_2007}.

\subsection{Constraints at the boundaries. 
\label{sec:boundary}}   

At the axis, the boundary conditions to be imposed are:
\begin{equation}\label{eq:bcs_0}
\left\{\;
\begin{matrix}
  f &=& 1\\
\noalign{\vspace{2mm}}
  f^\prime &=& 0
  \end{matrix}\;\right\} \quad \hbox{at} \quad \xi=0\;. 
\end{equation}
The second condition in Eq.~(\ref{eq:bcs_0}) is the usual prescription to prevent
the gradient of the solution (i.e. the electric current in this case) from
being undefined at the axis. The conditions of 
Eq.~(\ref{eq:bcs_0}) suffice to specify
the problem, and, for arbitrary $K$ and $H$, the solutions of
Eq.~(\ref{eq:selfsimilar_0}) extend to infinity. Here, we want to find
solutions with compact support.  Fixing \hbox{$r=\Rfront$}
(i.e. \hbox{$\xi=1$}) as the outer edge of the solution, an extra boundary
condition must be specified there.  For continuity, we first impose
\begin{equation}\label{eq:boundary_condition_edge_0}
f(\xi=1) = 0 \;.
\end{equation}
Then, to find the value of the derivative at the outer boundary, we use the
condition that the integral of the physical solution remains constant in
time; this makes sense at least in the ambipolar diffusion case, in which the
integral of the axial magnetic field, namely, the total magnetic flux $\Phi$,
is conserved. So:
\begin{equation}\label{eq:flux_self}
  \Phi = 2\,\pi\int_0^{\Rfront} \hskip -6pt r\,\difvar\,dr =
  2\,\pi\,\Bzero(t)\,\Rfrontsq(t) \int_0^1 \xi\,f(\xi)\,d\xi=\hbox{const}\;.
\end{equation}
In problems for which $\difvar$ does not change sign, this constant differs
from zero. This implies $\Bzero\,\Rfront^2 $ = const, so that, from
Eq.~(\ref{eq:B_R_constraint}), $K/H = 2$, which would directly lead to the ZKBP
solution. However, new self-similar solutions arise from imposing an
alternative possibility, namely,

\begin{equation}\label{eq:flux_balance}
\int_0^1 \xi\,f(\xi)\,d\xi=0\;,
\end{equation} 

\noindent for which $f(\xi)$ must pass through one or more nulls.  A
derivative condition at $\xi=1$ can then be obtained by integrating
Eq.~(\ref{eq:selfsimilar_0}) and applying Eq.~(\ref{eq:bcs_0}):
\begin{equation}\label{eq:integral_condition_2}
(K - 2\,H)\,\int_0^1 \xi\,f(\xi)\,d\xi = (-f^2 f')|^{}_{\xi=1}\;,
\end{equation}
from which
\begin{equation}\label{eq:boundary_condition_edge_1}
(f^3)' \to 0 \quad\hbox{as}\quad \xi \to 1\;.
\end{equation}
This condition is non-trivial, since the spatial self-similar profile can
have an infinite slope at the outer boundary. Physically speaking, this is
equivalent to saying that the diffusive flux vanishes at the outer edge; that is to say that the magnetic flux remains confined in the (expanding) domain of
the solution. To summarise the boundary conditions at the outer edge of the
solution, we have:

\begin{equation}\label{eq:bcs_d}
\left\{\;
\begin{matrix}
  f &=& 0\\
\noalign{\vspace{2mm}}
  (f^3)^\prime &=& 0
  \end{matrix}\;\right\} \quad \hbox{at} \quad \xi=1\;. 
\end{equation}

\subsection{Integration of the equations: usage of the phase plane}

Equation (\ref{eq:selfsimilar_0}) is an ordinary differential equation for
which standard numerical 
techniques (such as a second-order extended Euler method or a fourth-order
Runge-Kutta scheme) work well almost everywhere. However, our solutions have
infinite slope wherever $f=0$; the numerical integration can therefore be
inaccurate even for small steps of $\xi$ near such nulls.  We have, however,
devised a change of variable that renders the equation non-singular at the
nulls, so that the numerical procedure becomes smooth everywhere: when
nearing the nulls one can change the independent variable from $\xi$ to $f$
(admissible wherever $f' \ne 0$), use the diffusive flux $\difffluxnodim$ as
auxiliary variable, and split Eq.~(\ref{eq:selfsimilar_0}) into two
first-order equations:
\begin{equation}\label{eq:system_fdiff_of_f}
\hspace{3mm}\left\{\begin{matrix} \tderdps{\difffluxnodim}{f} &=& -
\fracdps{f^3\,K}{\difffluxnodim} \;+\; \xi\,H +
\fracdps{f^2}{\xi}\;,\\ \noalign{\vskip 3mm} \tderdps{\xi}{f} &=&
-\fracdps{f^2}{\difffluxnodim}\;.\hbox to 2cm{\hfill}\hfill
\end{matrix}\right.
\end{equation}
From Eq.~(\ref{eq:system_fdiff_of_f}), at an arbitrary null located at
$\xi_n$, the slope of the $\difffluxnodim(f)$ function reaches a finite
value, $\xi_n\,H$, which proves the adequacy of (\ref{eq:system_fdiff_of_f})
to carry the numerical integration through the nulls. In turn, however,
Eq.~(\ref{eq:system_fdiff_of_f}) becomes singular at the extrema of $f$,
where $\difffluxnodim=0$. Hence, when calculating solutions, the procedure to
follow is to switch from Eq.~(\ref{eq:selfsimilar_0}) to
Eq.~(\ref{eq:system_fdiff_of_f}), or viceversa, at intermediate locations
between each extremum and node (chosen not too close to the singular points
or nodes). Using this strategy, the numerical integration becomes smooth.

\section{Solutions}\label{sec:solutions_section}

\subsection{The eigenvalue problem}\label{sec:eigenvalue_problem} 

\begin{figure}[ht]
\includegraphics[width=\columnwidth]{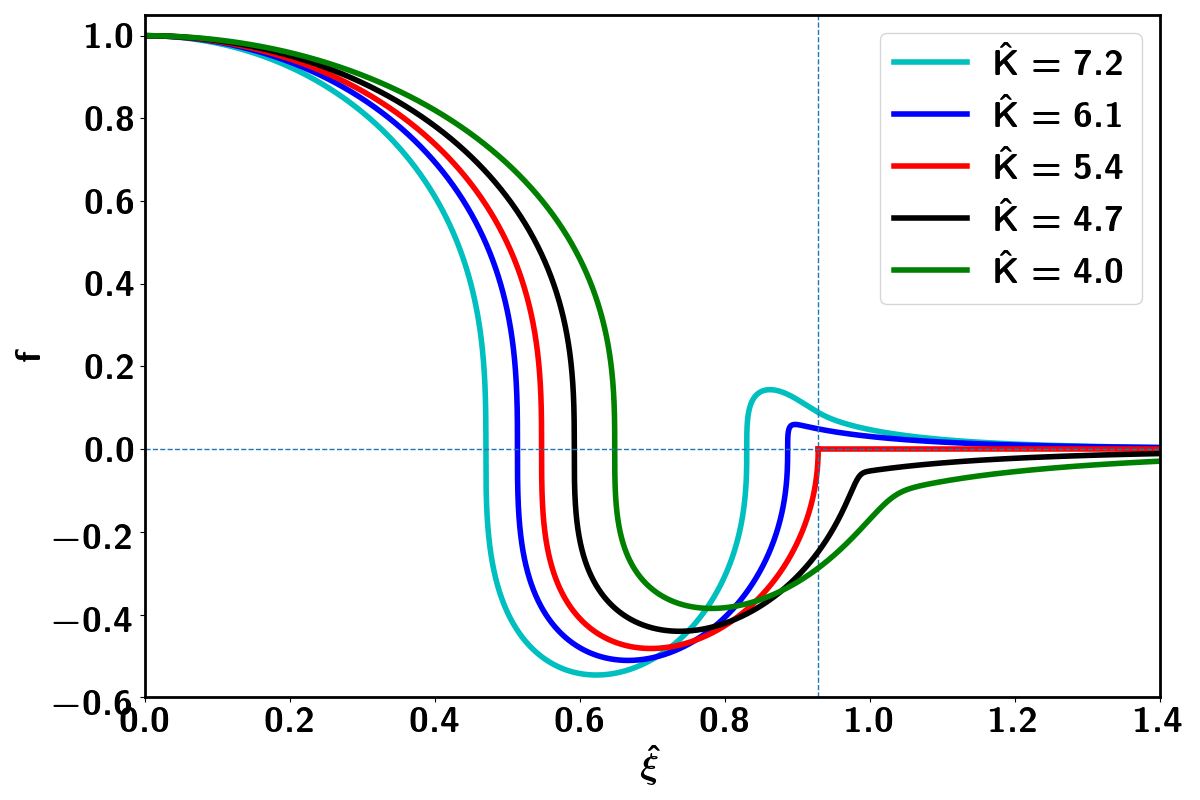}
\caption{\label{fig:across_eigenvalue} Selected solutions of the fundamental
  equation in an infinite domain subject to the inner boundary conditions
  of Eq.~(\ref{eq:bcs_0}) with asymptotic shape $f \propto \xi^{-\Ktilde}$ toward
  infinity that illustrate the identification of the eigenvalues and
  eigenfunctions for the fundamental equation subject to the full
    boundary conditions of Eqs.~(\ref{eq:bcs_0}),
    (\ref{eq:boundary_condition_edge_0}), and
    (\ref{eq:boundary_condition_edge_1}).}
\end{figure}

The solutions of Eq.~(\ref{eq:selfsimilar_0}) subject to the inner
boundary conditions of Eq.~(\ref{eq:bcs_0}) for arbitrary positive 
values $(K,H)$ in general extend to infinity, asymptotically behaving like $f
\propto \xi^{-K/H}$ (since the $f^2$ term decreases with $\xi$ much more
rapidly than the others).  To understand how to extract solutions that also
fulfil the outer boundary condition of Eqs.~(\ref{eq:boundary_condition_edge_0}) and
(\ref{eq:boundary_condition_edge_1}) at a finite radius $\xi=1$, we change the
independent variable to $\xitilde \,\equiv H^{1/2}\,\xi$, so that
Eq.~(\ref{eq:selfsimilar_0}) becomes
\begin{equation}\label{eq:selfsimilar_1}
  0\;=\; \Ktilde \,f \,+\, \xitilde\,f^\prime \,+\,
  \fracdps{1}{\xitilde}\left( \xitilde\,f^2\,f^\prime\right)^\prime \,,\quad
  \Ktilde\ \defdef\ \fracdps{K}{H}\;,
\end{equation}
the prime now understood as the derivative with respect to $\xitilde$.
Fig~~\ref{fig:across_eigenvalue} shows solutions for Eq.~(\ref{eq:selfsimilar_1}) for five values of $\Ktilde$ in the interval
$[4.0,7.2]$ that are suitably chosen so as to illustrate the behaviour of the
  solutions. Starting from $\xitilde=0$, all curves have an oscillatory
range, with one or two crossings of the horizontal axis, together
with an asymptotic tail extending to infinity. The change from one regime to
the other is quite abrupt, apparent almost as a corner when it occurs for
small $f$, that is, for $\Ktilde$ near the value for which the
asymptotic tail goes from below to above the axis.

In fact, for $\Ktilde = K_e \equiv 5.424$ (red curve), the asymptotic tail is
exactly zero and the end of the oscillatory range occurs at a precise point
($\xitildejunction=0.93$, dashed vertical line) with an infinite slope.  At the
junction, the condition $(f^3)'=0$ applies, since $(f^3)'$ changes sign when
varying $\Ktilde$ across the critical $\Ktilde$.  The solution for the
critical $\Ktilde$ therefore obeys the equation and both the inner and outer
boundary conditions, thus constituting an eigenfunction with just one
internal crossing of the axis (at $\xitilde=0.54$). Repeating this procedure
for solutions with more crossings, we find an infinite, but countable
collection of eigenfunctions which solve the full problem (as
  predicted by \citealt{Hulshof_1991}).  To transform the 
eigenfunctions back to unstretched coordinates, we set $\xi =
\xitilde/\xitildejunction$, $H = \xitilde_{r}^{\ 2}$, $\ K = \Ktilde_e
\,\xitilde_{r}^{\ 2}$, so that these solutions fulfil the original
Eq.~(\ref{eq:selfsimilar_0}) and all boundary conditions of
Eqs.~(\ref{eq:bcs_0}),  
(\ref{eq:boundary_condition_edge_0}), (\ref{eq:boundary_condition_edge_1}).

\begin{figure}[ht]
\centerline{\includegraphics[width=\columnwidth]{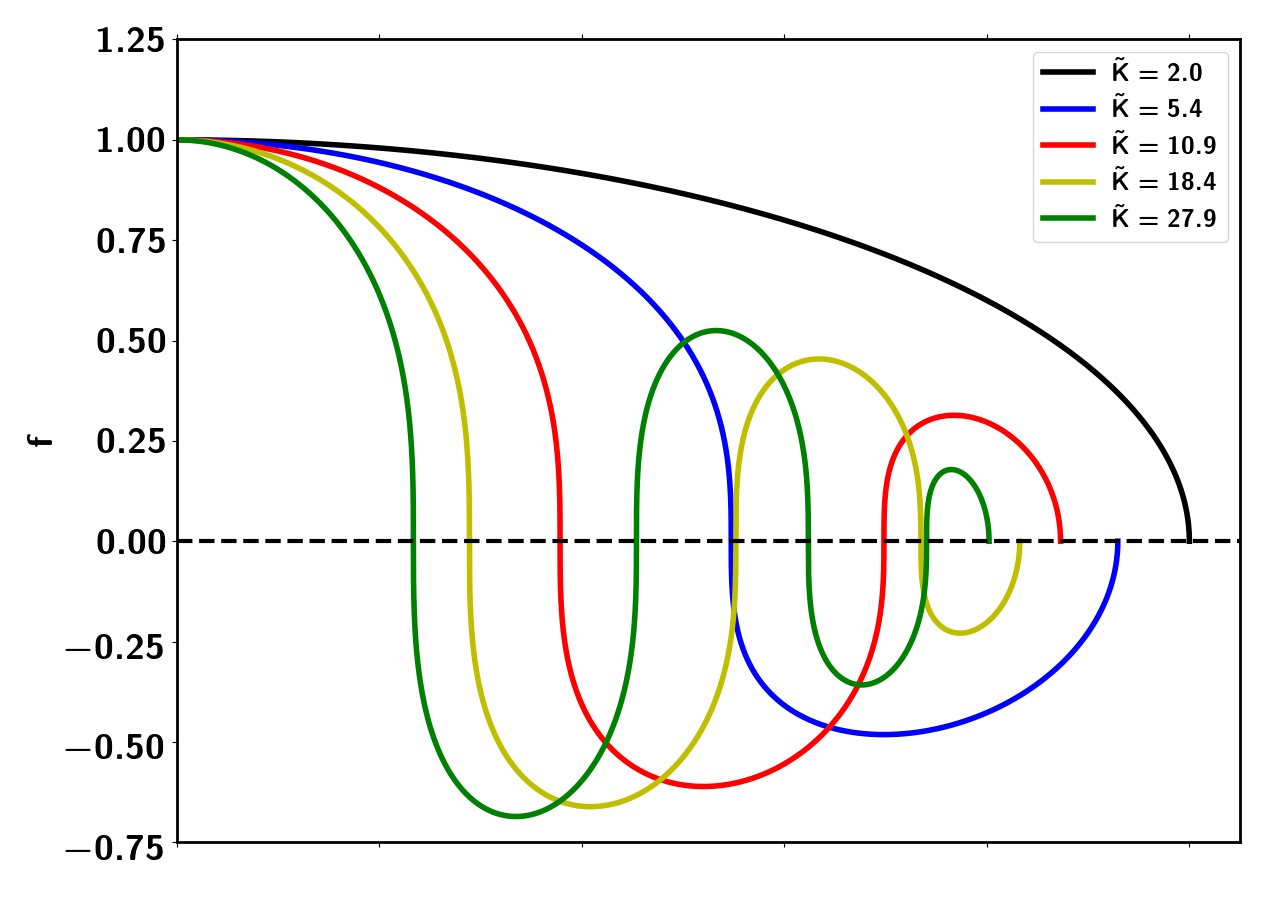}}
\centerline{
  \includegraphics[width=0.985\columnwidth]{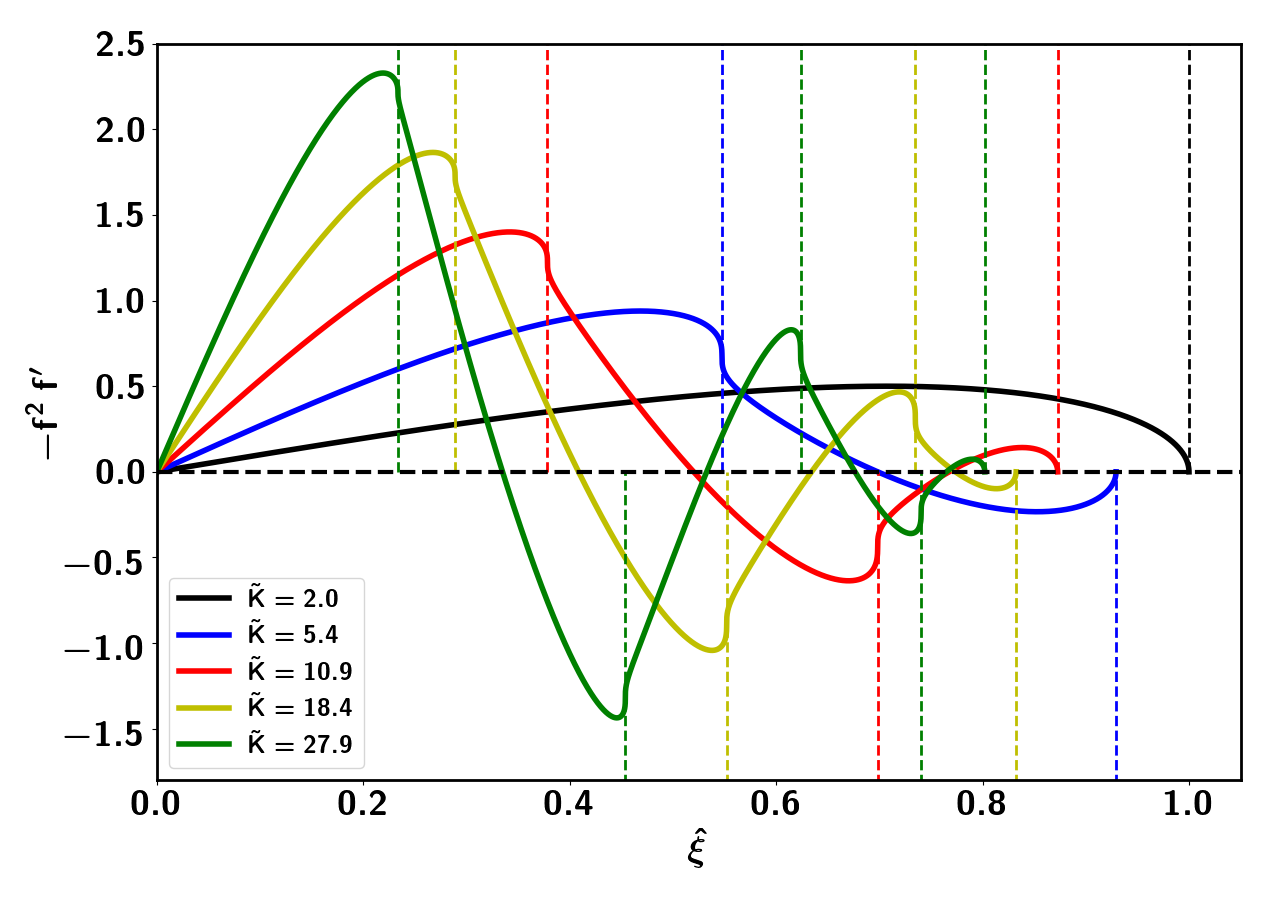}}
   \caption{\label{fig:the_solutions}  First five
     eigenfunctions as a function of the stretched coordinate $\xitilde =
     H^{1/2}\,\xi $.   The actual solutions $f(\xitilde)$ are shown in the
     upper panel, while the corresponding dimensionless diffusive flux
     $\difffluxnodim$ is displayed in the lower panel. The locations of the null
     crossings of 
 the solutions are indicated with vertical dashed lines.  In both panels, the
 eigenvalues $\Ktilde=K/H$ are indicated.  }
 \end{figure}

\subsection{The eigenfunctions\label{sec:solutions_subsection}}

Figure~\ref{fig:the_solutions} (top panel) presents the solutions of the
eigenvalue problem for the fundamental and the first four
harmonics; the fundamental eigenfunction is the ZKBP solution, which
  has the following simple profile:
\begin{equation}\label{eq:ZKBP}
f_{ZKBP}^{}(\xitilde) = (1-\xitilde^2)^{1/2}\;.
\end{equation}
In the bottom panel, the behaviour of $-f^2 f'$ (and so, of
  $\diffflux$) is presented for the five solutions shown in the upper panel,
  with the null crossings marked by vertical dashed lines.  Detailed runs of
$f$ and $-f^2\,f'$ as functions of $\xitilde$ for the first five harmonics is
provided in tabular form as attached files at a public repository
\footnote{\hbox{https://cloud.iac.es/index.php/s/pKZHJ2zPqYEWPfF}\label{ftn:2}}
  Table \ref{tab:one} provides the eigenvalues $\Ktilde$ and the locations of
  the end point, $\xitildejunction$, for those harmonics and the
    fundamental. For clarity in the diagram, the fifth harmonic is not shown
    in Fig.~\ref{fig:the_solutions}.

\begin{table}[b]
\caption{\label{tab:one} The first six eigenvalues ($\Ktilde = K/H$) together
  with the locations of the end points $\xitildejunction$ of the solution for
  the cylindrical case. The corresponding values of $K$ and $H$, and the
  exponents $\alpha$ and $\gamma$ of the power-laws (\ref{eq:B_R_constraint})
  -- (\ref{eq:exponents}) are also given. The actual run of the
    harmonics is provided as files available in the public repository
    indicated in footnote~\ref{ftn:2}. 
}
\vspace{3mm}
\centering
\begin{tabular}{||c||c|c|c|c|c|c||}  
\hline
\rule{0pt}{2.5ex}
\vtop to 3mm{\vfill}& $\Ktilde$  & $\xitildejunction$ & $H$ & $K$ & $\alpha$ & $\gamma$ \\
\hline
0&  $2.00$ &  1.000  & 1.000 &   2.00 & -1\hskip 1pt/\hskip 1pt3 &
1\hskip 1pt/\hskip 1pt 6 \\ \hline   
1&  $5.42$ &  0.929  & 0.864 &   4.69 & -0.422 & 0.0778 \\ \hline 
2&  $10.9$ &  0.873  & 0.762 &   8.30 & -0.458 & 0.0421 \\ \hline 
3&  $18.4$ &  0.833  & 0.693 &  12.8  & -0.474 & 0.0258 \\ \hline 
4&  $27.9$ &  0.802  & 0.644 &  18.0  & -0.483 & 0.0173 \\ \hline
5&  $39.5$ &  0.778  & 0.606 &  23.9  & -0.488 & 0.0123 \\ \hline
\end{tabular}
\end{table}

The functions resemble standard Sturm-Liouville solutions, except that the
crossings of the horizontal axis and the external boundary point have an infinite
slope. These singularities are essential, as can be seen by considering the
integral of the magnetic field in Eq.~(\ref{eq:flux_self}) restricted to the
interval between two successive internal nodes, ($\xi_n, \xi_{n+1}$), or
between the origin and the first node. Since the positions of the nodes are
fixed, the spatial part of the integral, $\int_{\xi_n}^{\xi_{n+1}}
\xi\,f(\xi)\,d\xi$, is constant. However, the product in front of the integral,
$\Bzero(t)\,\Rfrontsq(t)$, is a decreasing function of time, as easily proved
from Eq. (\ref{eq:B_R_constraint}) and noting that $\Ktilde$ increases upward of
$2$ as the harmonic number is increased. Now, when the diffusing physical
variable is a conserved quantity, the only possibility for its integral
between two nodes (or between the origin and the first node) to decrease in
absolute value is through diffusive flux to or from the neighbouring lobe:
$\difffluxnodim_n \,\defdef\, (-f^2\,f')_{\xi=\xi_n}^{} \ne 0\,$, so that
positive and negative values from neighbouring lobes mutually cancel.  The
need for a singularity at the nodes follows from that condition.

We can understand the nature of the singularities at internal nulls through
an asymptotic approximation. Choosing any of the nodes, say the one located
at $\xi_n$, and calling $\delta$ the distance from it, namely,
$\delta\,\defdef\,\xi-\xi_n$, we can show that close to the node
(more precisely: for $|\delta|/\xi_n \ll 1$, $|f K| \ll | \xi_n\,H f'|$ and
$f^2 \ll H \xi_n^2$), the following approximate solution is valid:
\begin{equation}\label{eq:current_sheet_solution}
  \delta \;=\; \fracdps{\difffluxnodim_n^2}{H^3\,\xi_n^3}
  \left[\fracdps{\Psivar}{2}(2-\Psivar) \,-\, \log(1+\Psivar)\right]\;,
\end{equation}
 with $\Psivar\,\defdef f\,H\,\xi_n/ \difffluxnodim_n\,.$
This implies that $|f| \propto |\hskip 1pt \delta\hskip 1pt |^{1/3}$ near the
node.  At the outer edge, the solution also has infinite slope, but the power
law is different, since there the diffusive flux vanishes. Indeed, it is
found that on the inside of (and near) the edge: $|f|~\approx~|\hskip 1pt 2\,
H\, \delta\hskip 1pt|^{1/2}$.

Some of the properties of the diffusive flux at the null crossings 
are clearly reflected in the curves shown in the bottom
panel of Fig~~\ref{fig:the_solutions}. For instance, the diffusive flux at those locations is different from zero, as
expected.  Secondly, the singularities at the nulls are apparent through the
vertical slopes of the curves.  Finally, the diffusive flux vanishes at the
outer edge, as it must to fulfil the boundary condition.

\section{The time-dependent solutions as tests of multidimensional MHD
  codes}\label{sec:tests}

\subsection{Scope of the tests}

The fundamental solutions of the self-similar problem presented in the
previous section can be used as basic tests to check the
correctness and capabilities of modules designed to solve the
ambipolar diffusion problem in general magnetohydrodynamics codes. Given the
presence of essential singularities within their domain, and the important
physical role played by the latter in the ambipolar diffusion
equation, these solutions can provide relevant, and demanding, tests that go
beyond the case used in the past, namely the ZKBP
solution, which has no internal nulls.
To show the potential of these new tests, in the following, we
  present examples of application to a specific numerical code, namely the
  Bifrost code. First, the code is briefly introduced
  (Sect.~\ref{sec:Bifrost_STS}). Then, a number of tests
are carried out (Sect.~\ref{sec:tests_harmonics}).  Finally, the asymptotic properties of the solutions
for large times are discussed (Sect.~\ref{sec:asymptotics}).

\subsection{The Bifrost code and the ambipolar diffusion module}
\label{sec:Bifrost_STS}

The numerical experiments are performed in double precision with the 3D
radiation-MHD Bifrost code \citep{Gudiksen_etal_2011}, whose core uses a
staggered mesh in cartesian coordinates and a sixth-order differential
operator. The code was used in a minimal configuration in which only the
induction equation with ambipolar diffusion term is solved (i.e. no mass,
motion, or energy equations are solved) in two dimensions $(x,y)$, imposing
cylindrical symmetry in the initial condition only.  For the ambipolar
diffusion term, we used the module developed by
\cite{Nobrega-Siverio_etal_2020b}, which employs the super time-stepping
(STS) technique \citep{Alexiades1996}. This module also includes an automatic
selection of the free parameters of the STS method to obtain the best
performance within a range that does not compromise the precision nor the STS
stability.  In addition, hyperdiffusion terms are included to guarantee the
stability of the code in regions with strong gradients. The coefficients for
the hyperdiffusion terms are chosen to be as small as possible, so that the stability
is preserved while minimizing the diffusion impact in the numerical
solution. The details of the automatic selection of the STS parameters and of
the hyperdiffusion coefficients can be found in Sects.~5 and 6 of the paper
by \cite{Nobrega-Siverio_etal_2020b}.  In that paper, the validation of the
module was achieved by reproducing the ZKBP solution in a situation with
cylindrical symmetry. The decay in time of the maximum field and the spatial
expansion of the outer front were found to follow an approximate power law,
with the exponents matching with $10^{-3}$ accuracy those of the self-similar
solution in Eq.~(\ref{eq:B_R_constraint}). Thanks to the STS implementation, a
speed-up factor of~8 compared with the simple implementation based on the
Courant-Friedrichs-Lewy criterion for the ZKBP test could be achieved.

\begin{figure}[h]
\centerline{\includegraphics[width=.5\textwidth]{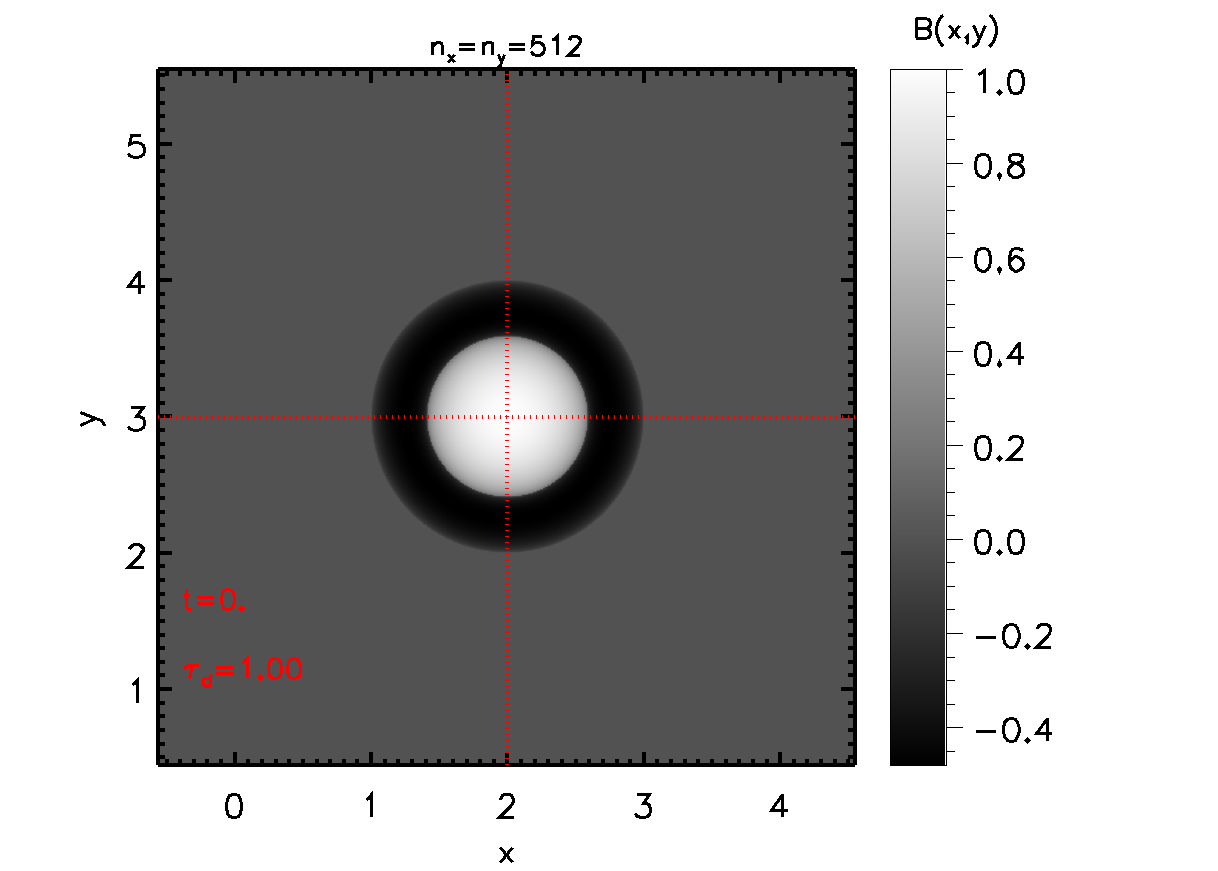}}
\centerline{\includegraphics[width=.5\textwidth]{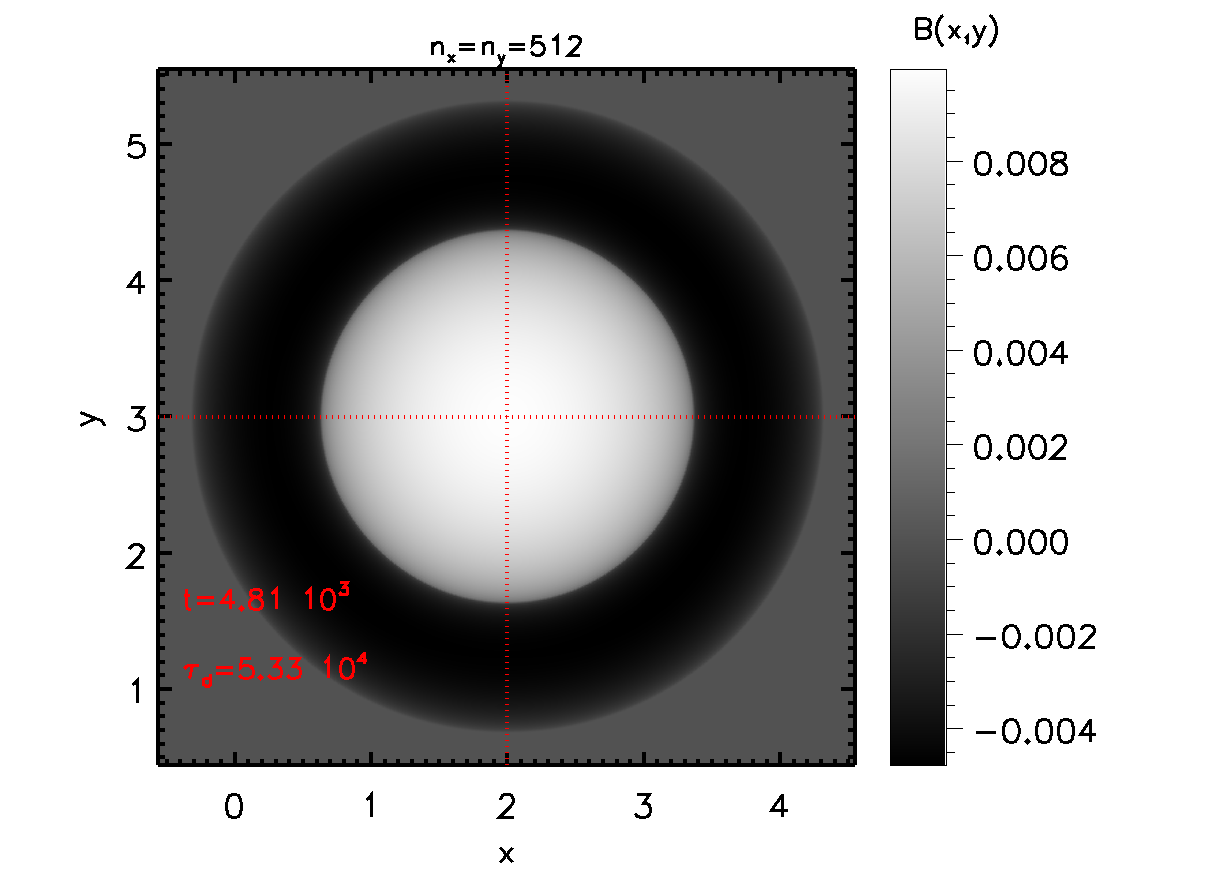}}
\caption{2D maps $B(x,y)$ showing the result of the
    Bifrost runs for the first harmonic at the initial time ($t=0$;
    $\tau_d=1$; upper panel) and at the end of the run ($t=\pot{4.8}{3};
    \tau_d=\pot{5.3}{4}$; lower panel). In the figure, the results 
    for the case with the lowest resolution, $n_x=n_y=512$, are presented. The
    decrease of the field strength with time can be clearly appreciated
    through the colour bar scales. 
  \label{fig:2D}}
\end{figure}

\begin{figure*}[h]
\centerline{
  \includegraphics[width=\textwidth]{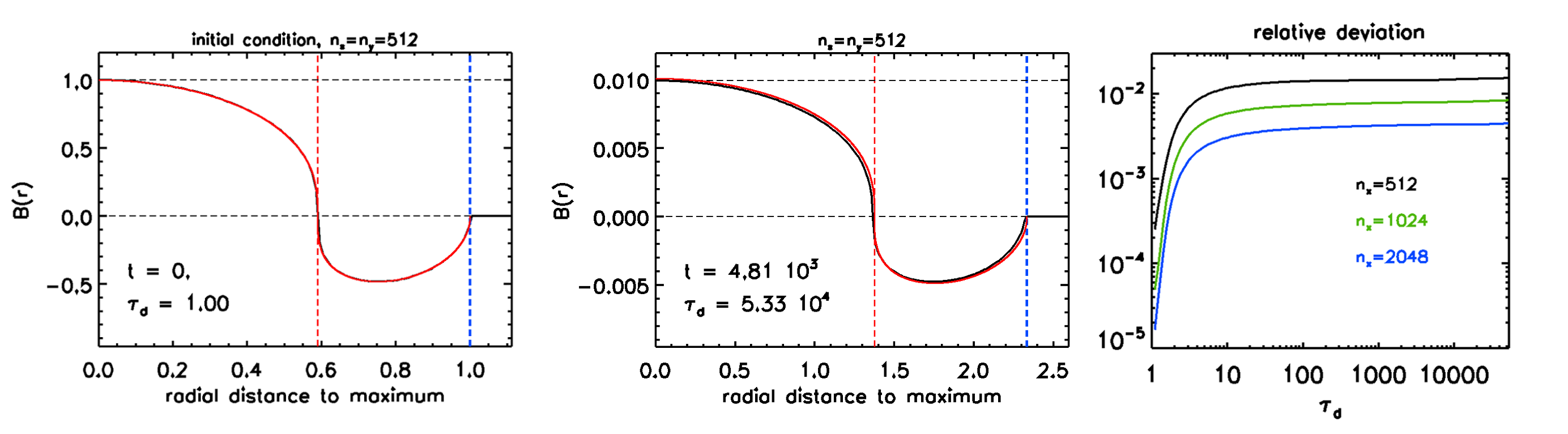}}
\caption{Illustration of the result of the Bifrost runs for the first
  harmonic. Left panel: Theoretical (red) and numerical (black) initial
  condition for the lowest-resolution case, $n_x=512$.  The positions of the
  outer front and of the singular current sheet for the numerical solution are
  marked with dashed blue and red lines, respectively.  Two horizontal lines
  are also superimposed at $B(r)=0$ and at the maximum of the field of the
  numerical curve.  Middle panel: Configuration at diffusive time
  $\tau_d=\pot{5.3}{4}$ (the full time evolution is presented in the
  accompanying movie, Animation~1).  Right panel: Relative mutual deviation
  of the maxima of the theoretical and Bifrost solutions for the three
  resolution levels used in the test.
\label{fig:bifrost_cylindrical}}
\end{figure*}

\subsection{Tests for different harmonics}\label{sec:tests_harmonics}

\subsubsection{First harmonic}\label{sec:tests_cylindrical_first}

To test Bifrost's STS module in the simplest configuration, we start by
focusing on the first harmonic discussed in
Sect.~\ref{sec:solutions_subsection}, which is shown as the blue curve of
Fig~~\ref{fig:the_solutions}~\footnote{In the following, we will refer to the solutions presented in the earlier sections of the paper as {\it theoretical solutions}, to distinguish them in the text from the solutions calculated with the Bifrost code, which we shall generally refer
  to as {\it numerical solutions}.}. 

For the test, we used the Bifrost code in two spatial dimensions $(x,y)$, with
the magnetic field pointing in the $z$-direction; the integration domain 
is a square of side $5.12$ units. 
As an initial condition, we 
used the theoretical solution for that harmonic centred at the centre of the
square; the outer front of the solution is located at a distance of unity
from the centre, which implies \hbox{$~\Rfront(t=0)~=~1$}, so as to fulfil
the normalisation condition in a simple way. Beyond the outer front, the
domain is padded with zeros. Three
levels of spatial resolution are used, namely $n_x = n_y = 512$,
$n_x=n_y=1024$, and $n_x=n_y=2048$, with grid sizes of $0.01$, $0.005,$ and $0.0025$,
respectively. The calculation is carried out until the maximum of the
solution is $10^{-2}$ times the initial value; this occurs at
$t=\pot{4.8}{3}$, which, using the values of $H$ and $K$ for the
first harmonic in Table~\ref{tab:one}, corresponds to the advanced diffusion
time $\tau_d = \pot{5.3}{4}$. 
Before any comparison is made with the
theoretical solutions of the previous section, we show in Fig.~\ref{fig:2D}
the $B(x,y,t)$ distribution at the initial (upper panel) and final (lower
panel) times of the run for the case with the lowest resolution
($n_x=n_y=512$). As shown in the figure, the symmetry of the initial 
condition is maintained to a high degree until the advanced time: a movie
containing the time evolution would just show equal expansion in all
directions. Simultaneously, from the field values of the grey-scale maps
as given in the colour bars, we see that the magnetic field is decreasing by a
large factor. The question, however, is whether the radial expansion and the
field decrease accurately follow the power laws seen in the past sections.

For the comparison of the Bifrost solutions with the theoretical
  ones, we 
first determine, at each timestep, the location of the maximum field, $[\xmax(t),
  \ymax(t)]$~\footnote{ According to the theoretical results, the maximum of
  the solution should not change location in the $(x,y)$ plane as time
  advances and, in fact, the actual numerical calculations discussed in the
  following fulfil this requirement, so we can drop the $t$-dependence in
  $(\xmax,\ymax)$. }. Then, the radial coordinate is defined as
\begin{equation}
r \;\defdef\; [(x-\xmax)^2+(y-\ymax)^2]^{1/2}\;.
\end{equation}
\noindent 
For the presentation of the results in the paper,  a zonal
averaging is carried out from the full 2D solution $B(x,y,t)$ at any given time, namely, a 1D radial distribution is obtained by
defining a collection of bins in the $r$-variable and averaging the value of
$B(x,y,t)$ for those $(x,y)$ whose radial distance to $(\xmax, \ymax)$ falls
within each given bin.
Concerning the theoretical
solutions, the time-dependence (\ref{eq:B_R_constraint}) -- (\ref{eq:tau_d})
and the scalings $\xitilde \,\equiv H^{1/2}\,\xi$ and $\xi = \xhat /
\Rfront(t)$
must be used with $H$ and $K$ coefficients given in the row
 corresponding to the first harmonic in Table~\ref{tab:one}
 (second row after the header).

In Fig~~\ref{fig:bifrost_cylindrical}, the initial function is shown in the
left panel; the final state at diffusive time $\tau_d=\pot{5.3}{4}$ for the
lowest-resolution case $n_x=n_y=512$ can be seen  in 
the middle panel.  The full time evolution for that case is presented in
an accompanying movie (Animation~1); in it, the
theoretical solution 
described in Sect.~\ref{sec:solutions_section} (red) and the Bifrost
solution (black) -- or, more precisely, the solution averaged over radial bins
of width equal to the grid size in the coordinate directions (i.e. $\Delta r
= 0.01$ for this case). For all animations in the paper, we used a snapshot
cadence 
corresponding to equal jumps in the logarithm of the diffusive time variable,
$\tau_d$.  The (numerically determined) position of the outer front is marked
with a dashed blue vertical line; 
that of the singular current sheet with a red dashed line. 
By construction, the theoretical and numerical curves lie on top of
each other at time $t=0$ 
(left panel). In the middle panel, we notice a small
mutual deviation at the final time.  To provide a quantitative measure for
the deviation which is independent of the binning used to draw the figures,
the absolute value of the relative 
deviation between the maxima of the theoretical and the Bifrost solutions for
each of the resolution levels is shown in the right panel as a function
of time.  The relative deviation first grows and then stays at a roughly
constant level in the advanced stages: this is due to the improvement of
spatial resolution that naturally comes about through the expansion of the
spatial support of the function as time progresses.

\begin{figure}[h!]
\hbox to \hsize{\hskip 2mm
\includegraphics[width=0.40\textwidth]{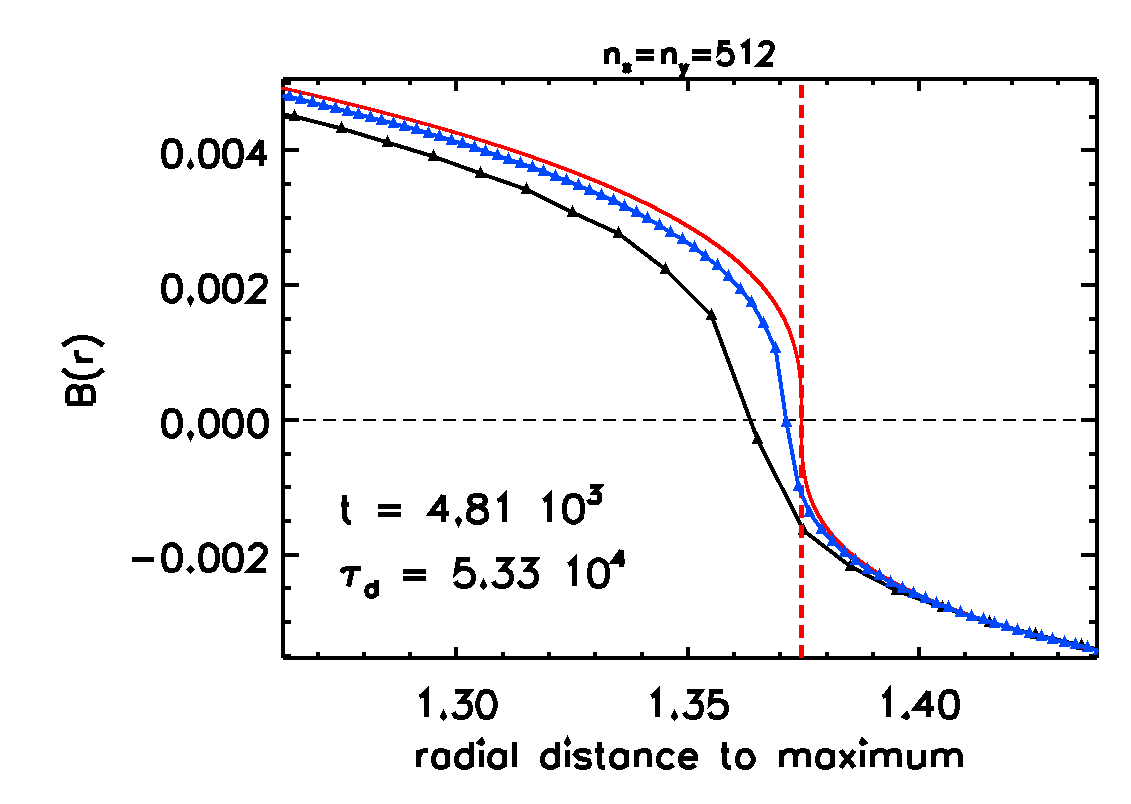}\hfill}
\caption{Solutions for the cylindrical problem in the immediate
  neighbourhood of the internal current sheet for the lowest- and
  highest-resolution cases ($n_x=n_y=512$, black, and $n_x=n_y=2048$, blue,
  respectively), in either case for the final time of the experiment. The red
  solid line is the theoretical curve.
\label{fig:bifrost_cylindrical_internal_current_sheet}}
\end{figure}

The general conclusion based on Fig~~\ref{fig:bifrost_cylindrical} and the
accompanying animation is that the match is excellent. Clearly, the most sensitive
location where one can check the mutual approximation between the theoretical
and numerical solutions is the internal current
sheet. Figure~\ref{fig:bifrost_cylindrical_internal_current_sheet} shows the
situation in the immediate neighbourhood of the sheet for the lowest- and
highest-resolution cases (black for $n_x=n_y=512$ and blue for
$n_x=n_y=2048$; the theoretical solution is drawn in red) and for the most
advanced time of the experiment. The locations of the centres of the radial
bins have been marked with symbols along the curves, for better
comparison. The three solutions are expanding outwards, but we can see that the
Bifrost solutions have a slight delay compared with the theoretical
one. Also, the case with highest resolution does a very good job at matching
the infinitely steep profile of the theoretical current sheet, whereas the
$n_x=n_y=512$ case has smoothed the profile to a certain extent.

\begin{figure*}[h]
\centerline{
  \includegraphics[width=1.02\textwidth]{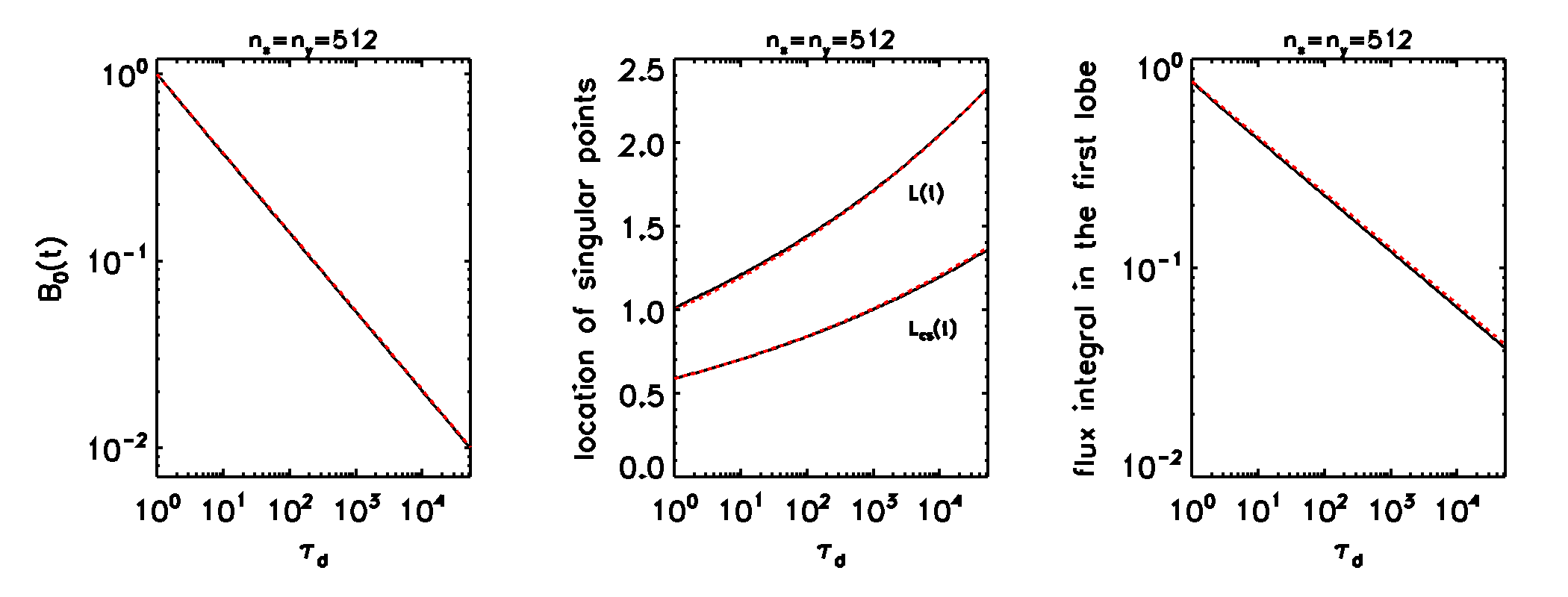}}
\caption{ Time evolution of various significant variables in the
    calculation for the first harmonic illustrated in
    Fig~~\ref{fig:bifrost_cylindrical}.  Left panel: Maximum of the
    field. Middle panel: Location of the singularities, with the upper curve
    corresponding to $\Rfront(t)$, and the lower one to $\Rcs(t)$. Right
  panel: Flux integral in the central lobe of the solution.  In each panel,
  the result of the Bifrost calculation is shown as a black solid line and
  the power laws derived from Eqs.~(\ref{eq:B_R_constraint}) -- (\ref{eq:tau_d})
  as red dashed lines.
\label{fig:power_laws_cylindrical}}
\end{figure*}

The quality of the Bifrost solution can also be tested by checking how
accurately it follows the theoretical power laws of decay and expansion
in Eqs.~(\ref{eq:B_R_constraint}) -- (\ref{eq:tau_d}). For the comparison we
  use the 
values of the exponents of those laws that correspond to the first harmonic,
namely $\alpha = -0.422$ and $\gamma = \pot{7.78}{-2}$ (Table~\ref{tab:one}).
The comparison is shown in Fig~~\ref{fig:power_laws_cylindrical} for the
worst-resolution case $n_x=n_y=512$ (black solid: Bifrost solution; red
dashed: theoretical laws). The left panel is for the maximum $\Bzero(t)$;
more demanding tests are provided by the positions of the outer front,
$\Rfront(t),$ and of the internal current sheet, $\Rcs(t)$, both shown in the
central panel. We note that the determination of the singular points for the
numerical curves in the central panel was done using the binned solution
explained in the previous paragraph.
The match is excellent: a minimum-square fit to the Bifrost curves yields
power-law exponents which match those of the theoretical law of
Eqs.~(\ref{eq:B_R_constraint}) -- (\ref{eq:tau_d}) with a precision of at 
  least a few significant digits; in relative terms, the approximation is
  $\pot{2}{-3}$ for $\Bzero(t)$ (left panel), $\pot{2}{-2}$ (middle panel,
$\Rfront$), and $\pot{1}{-2}$ (middle panel, $\Rcs$). The corresponding
precision for the highest resolution case $n_x=n_y=2048$ is $\pot{7}{-4}$ for
${\Bzero}(t)$, $\pot{5}{-3}$ for $\Rfront(t)$, and $\pot{4}{-3}$ for
$\Rcs(t)$, respectively. Of the three quantities, the location of the outer
front is the most difficult to determine, given the sharp corner between the
solution (which falls to zero as $|\,2\,H\,\delta\,|^{1/2}$, see
Sect.~\ref{sec:solutions_subsection}), and the zeros beyond the front. On
the other hand, the location of the internal current sheet is easier to
ascertain when one has the solution at hand, since it cuts the horizontal
axis with a steep slope. As apparent in the figure and in the match of the
power law exponents, both are well determined even in the worst-resolution
case. We also note that the precision increases roughly linearly
  with the number of grid cells, which reflects the first-order accuracy of
  the STS algorithm.

Throughout the paper we have pointed out that the different harmonics that
solve the eigenvalue problem associated with Eq.~(\ref{eq:selfsimilar_0}) provide a more
stringent test for MHD numerical codes than the basic ZKPB solution: the
reason is that the internal current sheets are non-avoidable singularities
but with finite diffusive flux across them.  This fact is essential for the
time evolution of the solution; the numerical calculation of the diffusive
flux is challenging precisely because of the infinite slope
of the theoretical solution at those points.  A further test of the capability of a
code to deal with this difficulty is obtained, therefore, by calculating the
integral of $B(x,y,t)$ in any of the lobes of the solution (i.e.
  calculating an integral such as Eq.~\ref{eq:flux_self}, but limited to a
  single cylindrical shell of constant magnetic field sign) and studying
its time dependence.  The integral of $B(x,y,t)$ in the ring between the
nulls or in the circle inside the internal current sheet must evolve in time
exactly as $\Bzero(t)\,\Rfront^2(t)$, and so, using the analytical
laws Eqs.~(\ref{eq:B_R_constraint})--(\ref{eq:tau_d}), as a power-law of
$\tau_d(t)$ with exponent $\alpha + 2 \gamma =
(2-\Ktilde)/[2(1+\Ktilde)]$. This implies (see Table~\ref{tab:one}) that the
integral is conserved in time for the ZKBP solution ($\Ktilde = 2$) and
decays increasingly rapidly for successively higher harmonics. For the first
harmonic, in particular, the power law is $\tau_d^{-0.267}$. A comparison
between numerical solution and analytical power law appears in the right
panel of Fig~~\ref{fig:power_laws_cylindrical}, again for the
worst-resolution case ($n_x=512$). The power-law exponent of the
minimum-square fit to the numerical curve matches the analytical value with a
accuracy of $\pot{8}{-3}$. For the highest-resolution case, $n_x=n_y=2048$,
the fit is four times more accurate, $\pot{2}{-3}$.
\vskip 5mm

\subsubsection{Higher harmonics}\label{sec:tests_third}

\begin{figure*}
\centerline{\includegraphics[width=\textwidth]{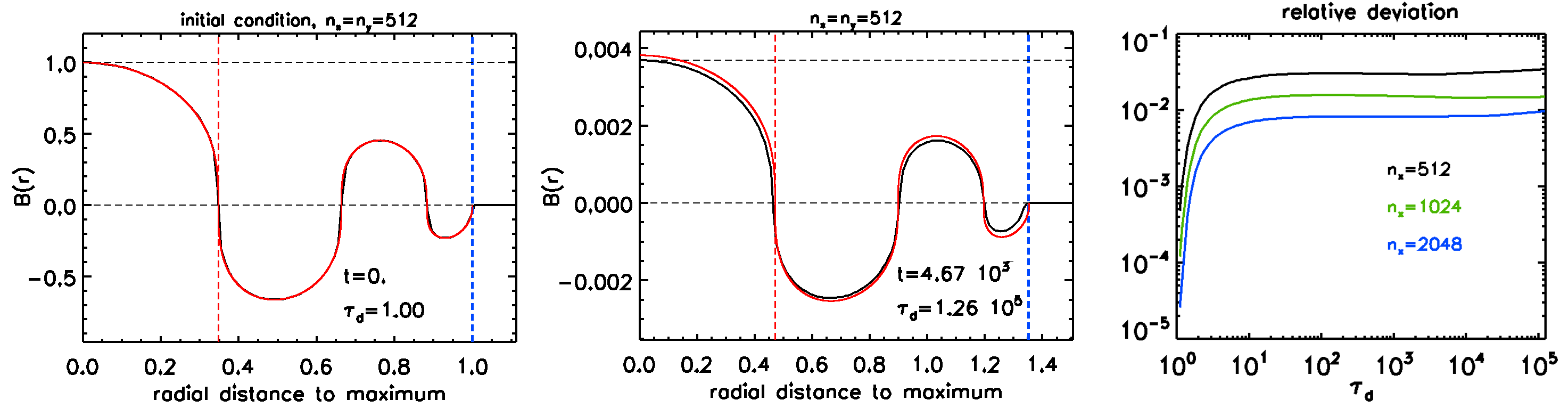}}
\caption{Illustration of the result of the Bifrost runs for the third
  harmonic. The left and middle panels show the initial condition and the
  final configuration in the calculation for $n_x=512$, respectively. The
  black thick curve is the numerical solution, whereas the red one is the
  theoretical solution. The position of the outer front and of the singular
  current sheet for the numerical solution are marked with dashed blue and
  red lines, respectively.  The full time evolution is presented in the
  accompanying movie, Animation 2. The right panel shows the relative mutual
  deviation of the maxima of the theoretical and Bifrost solutions for the
  three resolution levels used in the test.\label{fig:third_harmonic_main}}
\end{figure*}

To complete the section on the tests for various harmonics, we here
  study one 
with a higher number of null crossings.  As a relevant example, we consider
the third harmonic (light-green curve in Fig~~\ref{fig:the_solutions}),
which has a total of three internal null crossings in addition to the outer
front.  Checking with the theoretical solution, we expect (see
Table~\ref{tab:one}) a faster decay of the maximum in this case
($\alpha=-0.474$) and a much slower expansion of the spatial support
($\gamma = \pot{2.58}{-2}$). The result of the calculation for the
lower-resolution case ($n_x =n_y=512$) is presented in
Fig~~\ref{fig:third_harmonic_main} (left panel: initial condition; middle
panel: final snapshot in the series). The calculation was run until a similar
time $t$ as in the previous section; given the higher values of the sum $K+H$,
this implies a more advanced diffusion time $\tau_d(t),$ namely
 $\tau_d=\pot{1}{5}$, with the maximum of the solution
decreasing to $\pot{4}{-3}$ of the initial value.  We see that the numerical
solution deviates from the theoretical one more markedly than in the
corresponding case for the first harmonic
(Figure~\ref{fig:bifrost_cylindrical}): here, the effective resolution for
each of the lobes is roughly twice as bad as for the run for the first
harmonic. This is also reflected in the corresponding curves for the relative
deviation of the maxima of the solution between the Bifrost and theoretical
solutions (right panel of the figure): for the highest-resolution case ($n_x
= n_y = 2048$), the maximum relative deviation, while still quite good (on the
order of $\pot{}{-2}$), is a factor $2$ greater in the experiment for the
third~harmonic than in that for the first. For the low-resolution case, the
relative deviation is in the range of a few percent.

\begin{figure*}[ht]
\centerline{
  \includegraphics[width=1.0\textwidth]{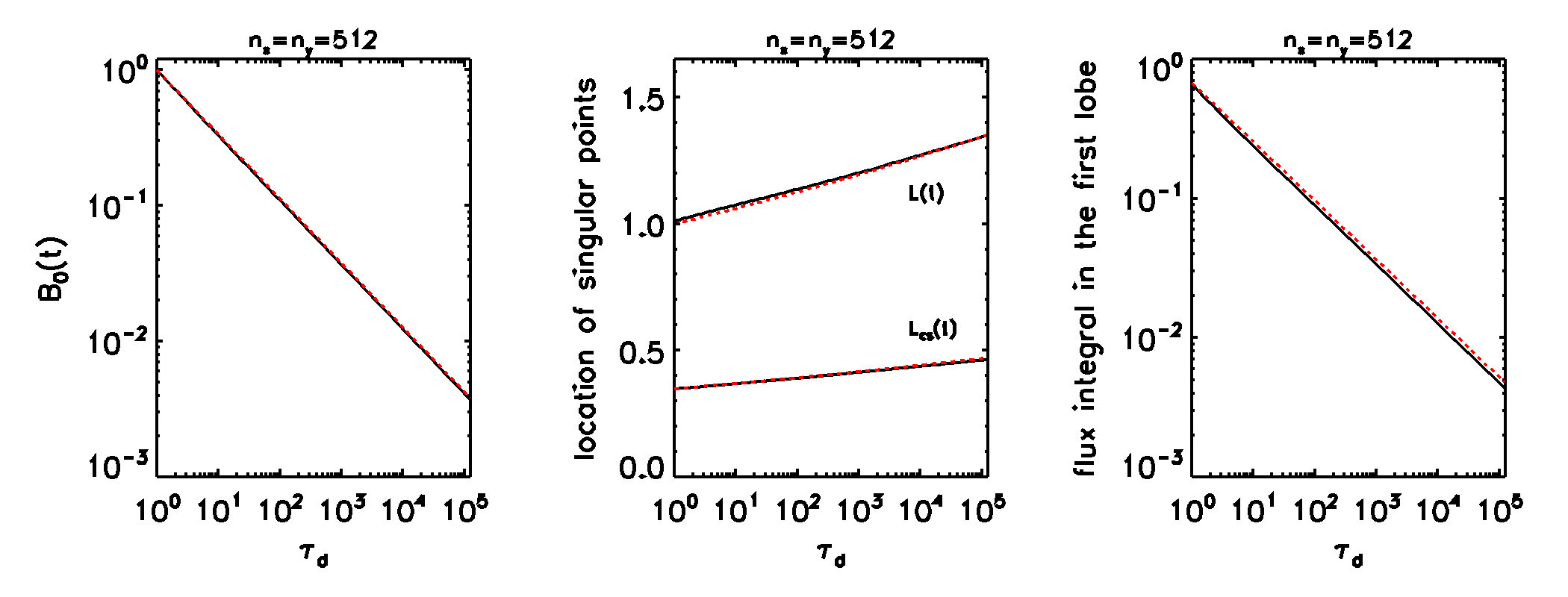}}
\caption{ Time evolution of various significant variables in the
    calculation for the third harmonic illustrated in
    Fig~~\ref{fig:third_harmonic_main}. Left panel: Maximum of the field.
    Middle panel: Location of the singular points, where the upper curve is
    for $\Rfront(t)$, and the lower one for $\Rcs(t)$. Right panel: Value of
    the flux integral in the central lobe of the solution. In each panel, the
    result of the Bifrost calculation is shown as a black solid line; the
    power laws derived from Eqs.~(\ref{eq:B_R_constraint}) -- (\ref{eq:tau_d}) are
    shown as red dashed lines.
\label{fig:power_laws_third_harmonic}}
\end{figure*}

The determination of the location of the singular features and of the
corresponding power-law exponents is a challenging test for the
lower-resolution case ($n_x=n_y=512$) of this harmonic or higher. The
location of the outer front $L(t)$ is calculated by the numerical solution
with a relative maximum deviation of $\pot{7}{-3}$ from the theoretical
value; the 
fit of $L(t)$ to a power law yields an exponent that matches the
theoretical value, $\gamma$, within $\pot{5}{-2}$. For the innermost
current sheet, the situation is more demanding for the numerical solution:
$L_{cs}^{}(t)$ is determined with accuracy better than $\pot{4}{-3}$
throughout the calculation; however, the fit of $L_{cs}(t)$ to a power law
yields an exponent that matches the theoretical value only within
$\pot{4}{-2}$. The fitted power laws for the time decay of the maximum of
the solution and of the integrated magnetic flux within the innermost lobe
have exponents that match the theoretical ones with an acceptable accuracy
of $\pot{3}{-3}$ in the first case and $\pot{9}{-3}$ in the second.  For
the higher-resolution cases the 
Bifrost calculation does a much better job. For the highest resolution case
($n_x=n_y=2048$), the determination of the power law exponents matches the
theoretical values with accuracy $\pot{1}{-3}$ (decay of the maximum),
$\pot{2}{-2}$ (location of the outer front), $\pot{1}{-2}$ (location of the
innermost current sheet) and $\pot{3}{-3}$ (flux integral).

\subsection{Tests for more general initial conditions: the asymptotic
evolution in time}{\label{sec:asymptotics}}

In the absence of perturbations, the pure eigenmodes must evolve exactly
fulfilling the self-similar shape and the power laws
of Eqs.~(\ref{eq:B_R_constraint}) -- (\ref{eq:tau_d}). However, given the errors
associated with the numerical solution of the equation (both those associated
with the discretisation of the initial condition and with the accumulated
error along the update in time), the numerical solution will always deviate
(if only by a small amount) from the pure theoretical eigenmode. However, in the
previous section we show that the match of the numerical solution to the
theoretical eigenmodes is excellent all the way to very advanced diffusion
times, such as $\tau_d = \hbox{O}(\pot{5}{4})$; this is a qualitative indication
both of the validity of the numerical solution and of the fact that the
eigenmodes of the problem themselves, when subjected to small perturbations,
remain in the close neighbourhood of the exact eigenmode up to advanced
diffusion times. 

A further question concerns the time evolution of the (theoretical
  and numerical) solutions when the initial condition deviates from a pure
  eigenmode by a finite perturbation:\ we consider whether the exact eigenmodes act as attractors for
  neighbouring functions -- at least for those with the same number of
  zero-crossings of the solution. This question is of interest from the point of
  view of applied mathematics, of the physics of ambipolar diffusion, and of
  the construction of AD modules in numerical codes. In this section, we first
  obtain hints from the Bifrost solutions of the problem for the first
  (Sect.~\ref{sec:arbitr_1st}) and higher (Sect.~\ref{sec:arbitr_3rd}) harmonics, which
  represent two different patterns of behaviour. Then we briefly discuss a
  few results from the literature (Sect.~\ref{sec:mathem}).

\begin{figure*}[ht]
\centerline{\hskip -1cm
  \includegraphics[width=0.53\textwidth]{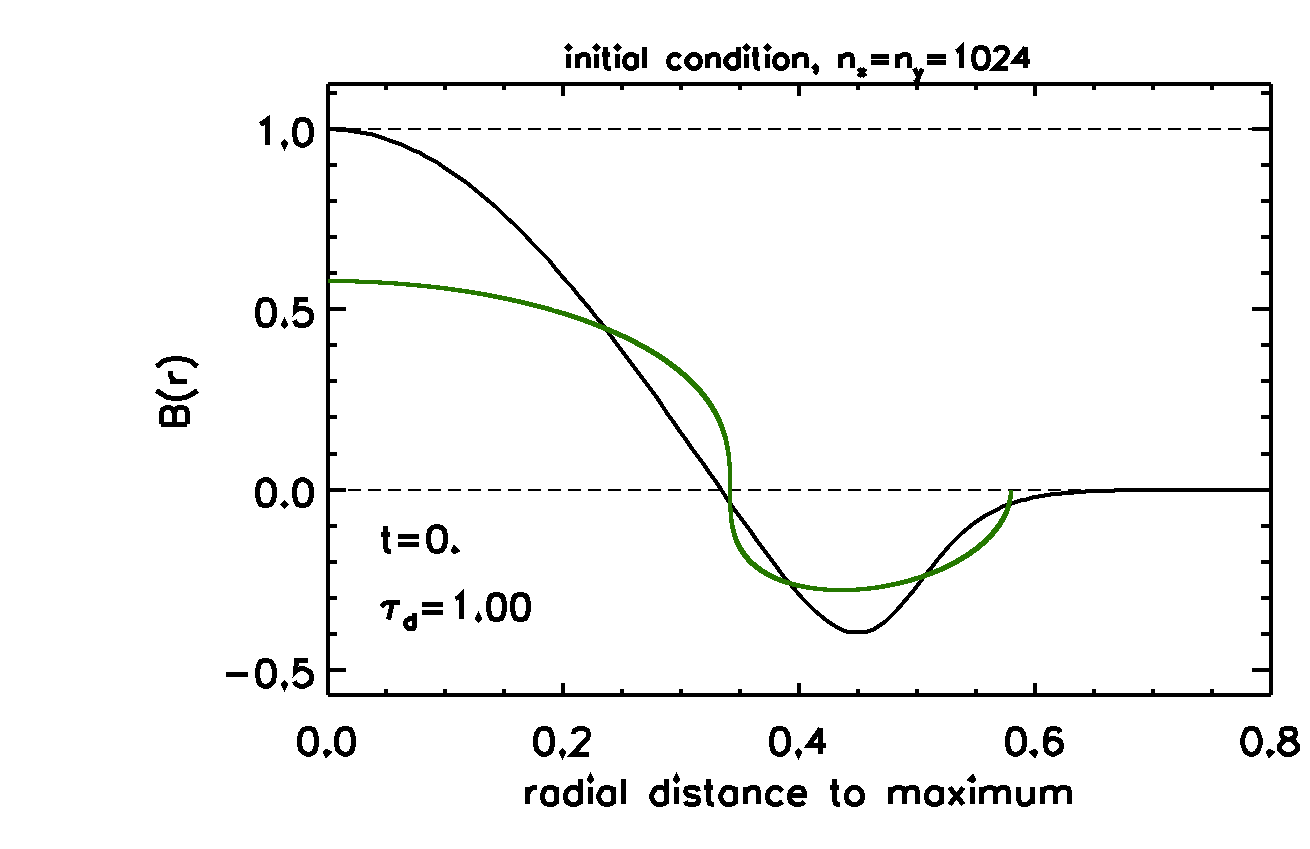}
  \hskip -1mm
  \includegraphics[width=0.53\textwidth]{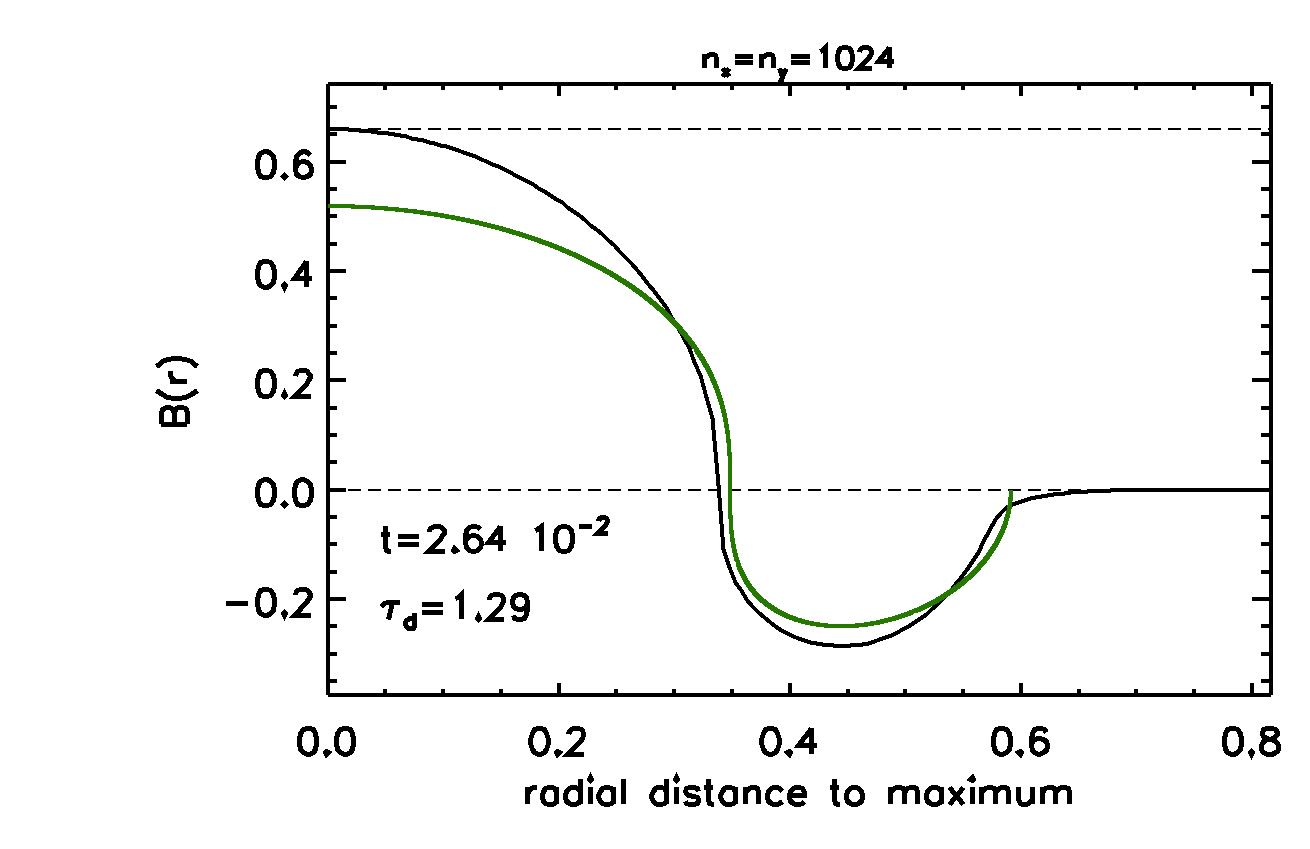}}
\vskip -0mm
\centerline{\hskip -1cm
  \includegraphics[width=0.53\textwidth]{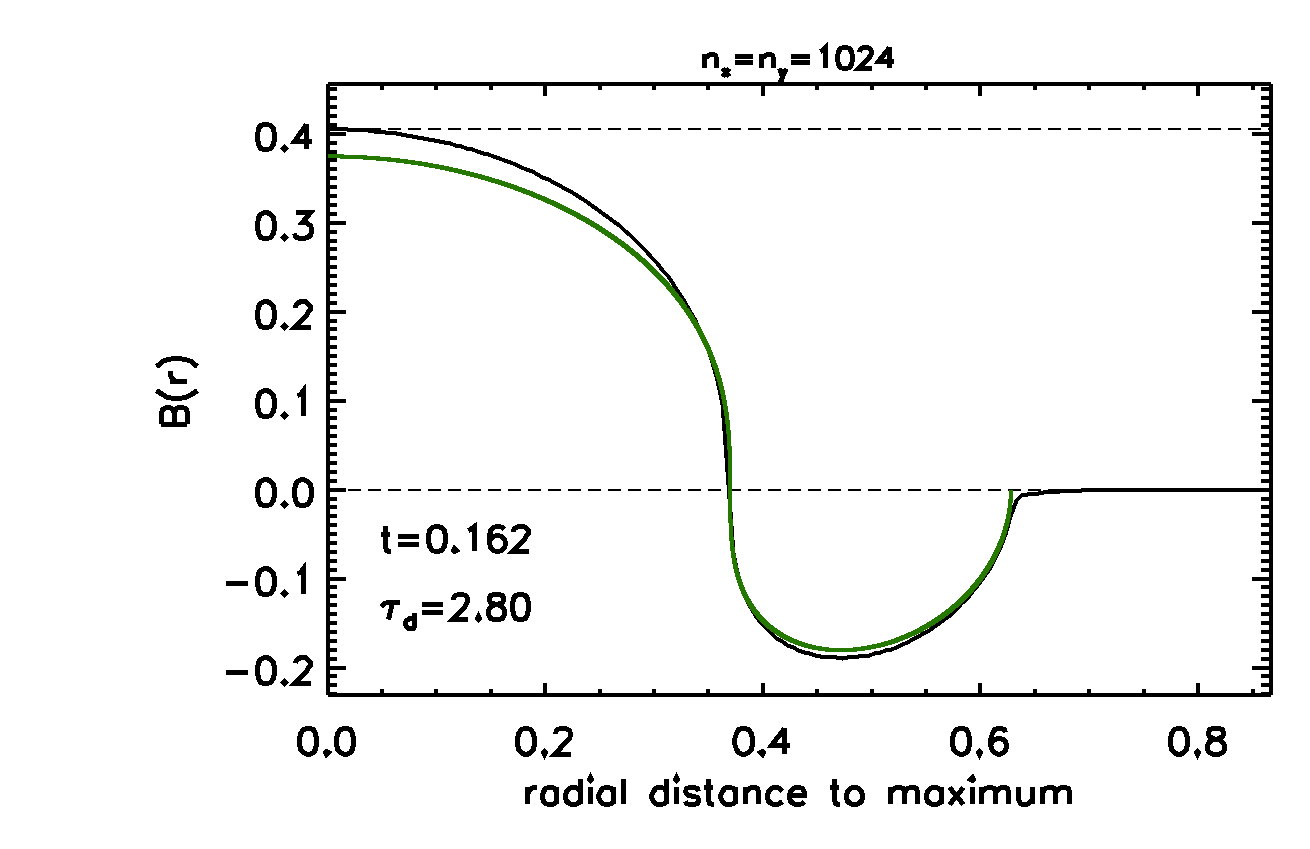}
  \hskip -1mm
  \includegraphics[width=0.53\textwidth]{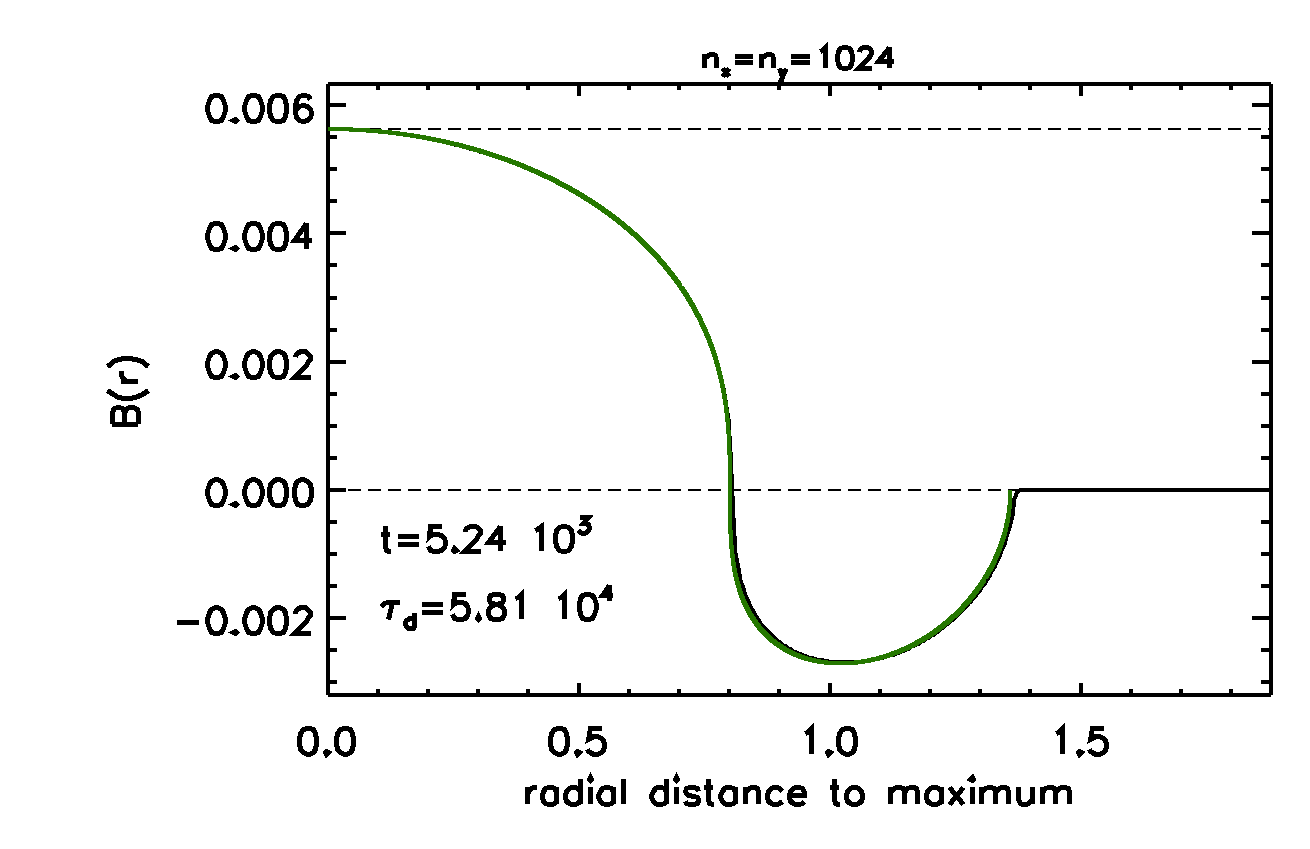}}
\caption{
Four snapshots corresponding to the time evolution 
for initial condition with zero net flux, compact
  support, and a single internal null point, but otherwise of arbitrary
  shape. These snapshots are taken from the accompanying
    Animation~3, where the full time evolution can be seen. In both the
    figure and animation, the shape of the first harmonic corresponding to
    the unsigned flux   of the solution toward the end of the evolution is superimposed as a green  curve. }  \label{fig:cylindrical_arbitrary}
\end{figure*}

\subsubsection{The first harmonic}\label{sec:arbitr_1st}

Figure~\ref{fig:cylindrical_arbitrary}, which shows a selection of
  four  snapshots of the accompanying Animation~3, presents
(black curves) the results of a test run with Bifrost for the first harmonic
in which the initial condition deviates by a significant amount from the pure
eigenmode. For that example, the initial condition is given by:

  \begin{equation}\label{eq:arbitr_1st_harm_ic}
\fracdps{1}{2}\cos\left(\fracdps{3\pi}{2}\,r\right)\;
  \left[1-\tanh\left(\fracdps{r-r_1}{W_1}\right) \right] \quad\hbox{in}
  \quad r\in[0,1]\;, 
\end{equation}
and zero outside of that interval. The parameters $r_1$ and $W_1$ were chosen
as $W_1=0.06$ and $r_1=0.48432,$ so that the function (top left panel) has
zero net magnetic flux, only one internal null and essentially zero
derivative at the origin; intermediate resolution ($n_x=n_y=1024$) was chosen
for this example. By letting the code run, we observe that the shape of the
solution quickly approaches that of the first eigenmode: we added to the
panels (green curves) a first harmonic calculated with the parameters
(unsigned flux and maximum field) which hold toward the end of the evolution:
we see that the initial solution quickly tends toward that shape; for
instance, at the comparatively early diffusive time $\tau_d=2.8$ (lower left
panel in the figure; see also the accompanying Animation~3) the solution is
already quite near the first harmonic toward which it is advancing
asymptotically. The run is carried out until a very advanced diffusive time,
$\tau_d = \pot{6.9}{4}$ (lower right panel); from about $\tau_d =
\hbox{O}(10^3)$ onward, the two curves are almost superimposed on each other,
the largest discrepancy being found near the outer front.

We ran further cases of finite perturbation to the first
  harmonic. In all cases, the solution quickly approaches the 
  shape of the exact first eigenfunction. This is highly suggestive that the
  first harmonic is not just stable against small perturbations (as shown in
  Sect.~\ref{sec:tests_cylindrical_first}) but also in the sense that it is
  an attractor at least for initial functions that have zero net flux and a
  single zero-crossing in the middle of the domain.

Of special importance is the condition of zero net flux: when running
  those cases one has to make sure that the initial condition is in flux
  balance, at least up to a few significant digits; in other words, the
  integrated net flux must be orders of magnitude below the unsigned
  flux. From the theory (see Sect.~\ref{sec:mathem}), we know that an initial
  condition which is not in flux balance will evolve into the simple ZKBP
  solution corresponding to the same net flux. However, this is not going to
  modify the solution to any major extent until it has evolved so much that
  the unsigned flux of its lobes is comparable to the net signed flux of the
  initial condition. With the integrated flux evolving as
  $\Bzero(t)\,\Rfront(t)^2$, that is, as $\tau_d^{\alpha+2\gamma}$, the
  evolution will need to be calculated up to large values of $\tau_d$ for the
  ZKBP form to begin to become apparent in the solution; for instance, if the
  initial flux imbalance is $10^{-4}$ in relative terms, $\tau_d$ ought to be
  enormously large, possibly on the order of $10^{13}$, for the ZKBP function to
  begin to dominate the shape of the solution.

\begin{figure*}[ht]
\centerline{
\includegraphics[width=1.0\textwidth]{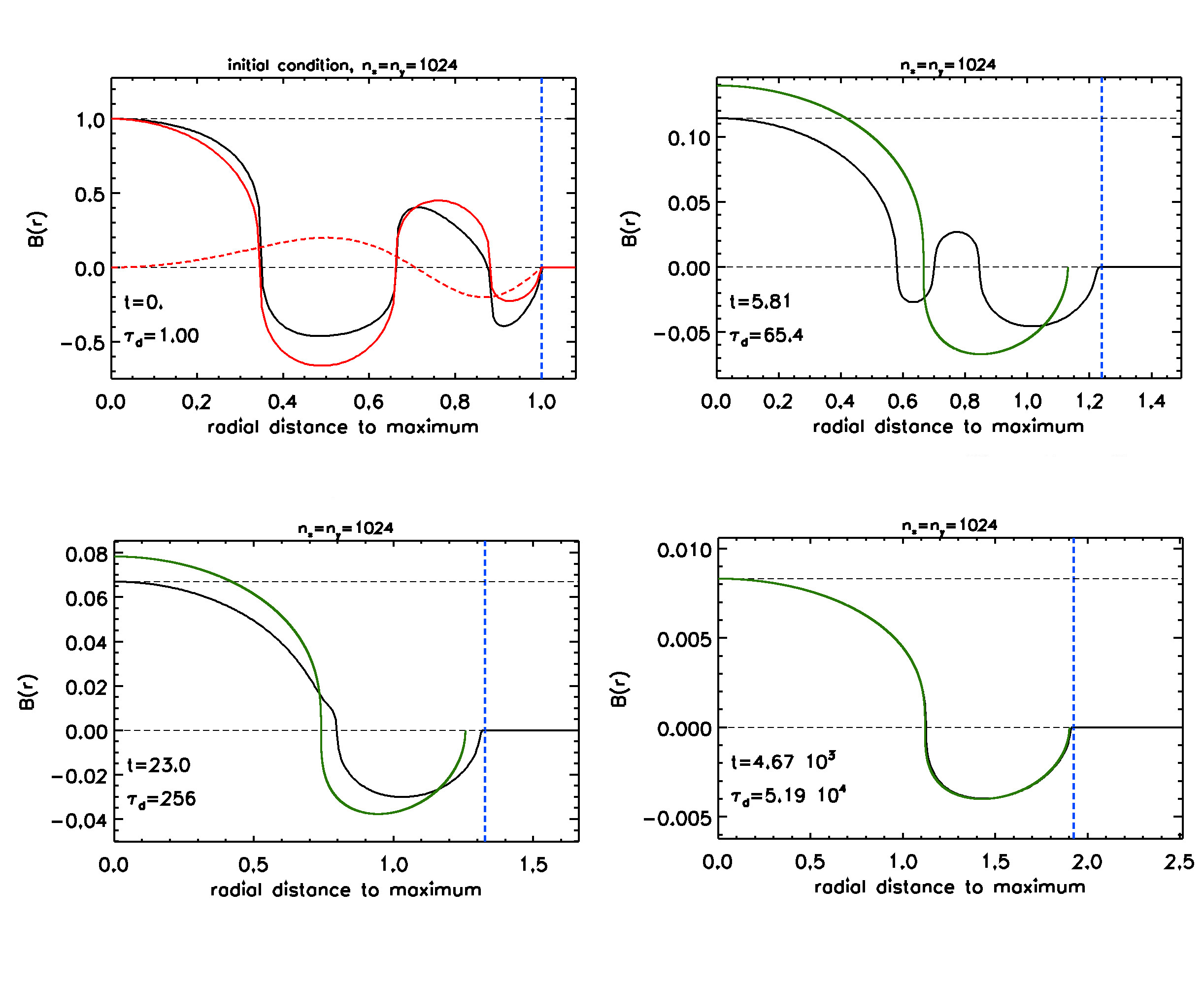}
}
\caption{Selected snapshots from the time evolution of the third harmonic
  when a non-small perturbation is added initially (see also Animation~4 for
  a high-cadence display of the evolution). The top left panel shows the
  initial condition (black solid), and its components, namely, an exact third
  harmonic (red solid) and a non-small perturbation (red dashed). The other
  three panels highlight the gradual conversion of the solution toward the
  shape of the first harmonic. The green curve corresponds to the time
  evolution of the first harmonic with unsigned flux and maximum field
  matching the solution in the final snapshot. The value of $\tau_d$ in the
  images is calculated with the $H$ and $K$ of the first harmonic
  (second row in Table~\ref{tab:one} after the header). The
  corresponding values with the $H$ and $K$ for the third harmonic would be:
  $\tau_d = 1$;\/ $157$;\/ $619$ and $\pot{1.26}{5}$, respectively.
\label{fig:arbitr_3rd_harm}}
\end{figure*}

\subsubsection{Higher harmonics}\label{sec:arbitr_3rd}

To test the behaviour of the higher harmonics when supplemented with a finite
perturbation, we reran an experiment for the third harmonic, but this time
adding a zero-flux perturbation such that the resulting function has the same
number of zero-crossings along the radial direction as the original
eigenmode, namely, three internal ones and the outer front.  The evolution in
this case is markedly different to all previous cases (see
Fig~~\ref{fig:arbitr_3rd_harm} and the accompanying Animation~4): the time
evolution leads to a transformation of the third harmonic through merging of
pairs of internal nullpoints~\footnote{These processes of
      disappearance of internal nulls have been discussed in a somewhat
      different context by \citet{Hulshof_etal_2001}.}.  In the
    figure and animation, the numerical solution is the black solid line.
    For the initial condition (top left panel in the figure) we additionally
    show its constituents, namely, a pure third eigenmode (red line) and the
    zero-flux, finite perturbation added to it, namely,
    $0.2\,\sin(2\,\pi\,r^2)$ (red dashed). 
As time proceeds, the shape of the solution increasingly deviates
  from that of a third harmonic: in the top-right panel, the two innermost
  nulls are about to merge, a process which is already complete at the time
  of the bottom-left panel. Finally, as time advances, the solution tends to
  the shape of the first harmonic with remarkable precision. For proper
  comparison, we have determined an exact first harmonic with the parameters
  (unsigned flux and maximum magnetic field) corresponding to the final state
  of the numerical solution and calculated its profile backward in time using
  the standard power-laws for that harmonic. That profile is superimposed (as
  a solid green curve) to all snapshots with $\tau_d > 10$. 

Finally, we also carried out a number of further experiments
  with different finite perturbations to the third harmonic, and also to the
  fifth harmonic. The results are similar in terms of null-point merging and
  transformation to the first harmonic after a finite time as in the
  experiment in this subsection.

The empirical results obtained here are compatible with the
  possibility that initial conditions consisting of a zero net-flux finite
  perturbation to harmonics higher than the first (and such that the
  superposition has compact support) end up with an exact first-harmonic
  shape. Also, judging by our results, it is not necessary to go to very
  large values of the diffusive time, $\tau_d$, for this process to be well
  advanced, thus constituting a case of what is known in the mathematical
  literature as 'intermediate asymptotics'. On the other hand, when
  using really poor spatial resolution to solve the problem of the evolution
  of the higher harmonics with no explicit finite perturbation, the
  numerical errors are likely to lead to a transformation
  of those harmonics into the first harmonic in a comparatively short integration 
  time. Therefore, testing the higher harmonics with an MHD code adds a layer
  of difficulty to the test: whenever the initial discretisation or the
  numerical error made during the update in time is not small, the higher
  harmonics will evolve toward the first one and the test will fail.

\subsubsection{The mathematical theory}\label{sec:mathem}

A number of fundamental results have been rigorously proved in the
  mathematical literature concerning the asymptotic behaviour in time of some
  of the solutions of the porous medium equation and related equations
  \citep[e.g.][]{Kamin_and_Vazquez_1991, Bernis_etal_1993,
    Hulshof_etal_2001}. What is of interest for us here is, primarily, the
  results 
  that can be applied to the cylindrically symmetric case with diffusion
  coefficient which is proportional to the square of the dependent variable
  ($n=2$, $m=3$ in the notation of
  Eq.~\ref{eq:porous_medium_equation}). The most basic result, already
  mentioned in Sect.~\ref{sec:arbitr_1st}, is that initial conditions which have
  a finite nonzero flux integral (called 'the mass' in the mathematical
  literature for the PME) converge toward the ZKBP solution with the same
  flux integral ('mass') asymptotically in time \citep[][Theorem
    18.2]{Vazquez_2007}; here, allowance is made for either a positive or
  negative flux integral by globally changing the sign of the ZKBP solution;
  also, 'convergence' is meant in the sense that the $L^p$ norm of the
  difference between the actual solution and the ZKBP function tends to zero
  as $t \to \infty$ faster than a negative power of the time with an exponent
  which is a function of $n$, $m,$ and $p$ (e.g. $-1/3$ for $n=2$ and $m=3$
  in the $L^2$~norm; see details in the book by \citealt{Vazquez_2007}).  A
  complementary result is the following: when the initial condition has
  positive net flux and its negative part has compact support, then the whole
  solution evolves into a positive function after a finite time
  \citep[][Theorem 18.29]{Vazquez_2007}.  Since we are dealing with signed
  functions which have zero flux integral, these results are of interest
  mainly because they impose a strict condition on the possible flux
  imbalance caused by numerical errors (as discussed in
  Sect.~\ref{sec:arbitr_1st}, final paragraph): if it is not small, the numerical
  solutions will approach the ZKBP solution in a comparatively short
  time. However, the flux imbalance in all the Bifrost experiments discussed
  in the present paper is small enough that they have not shown this behaviour
  even though they have been run until a very long diffusive time.

Of special interest in the present paper would be mathematical
  results concerning the first harmonic in the eigenfunction series, meaning
  the harmonic with a single zero crossing within its spatial support.
  \citet{Bernis_etal_1993} deal with the plane-parallel 1D problem; the paper
  is devoted to the study of the so-called 'dual porous medium equation'
    (DPME), which is obtained through a double integration in space of the
  standard PME. These authors prove the existence of  an asymptotic trend of
  the solutions of that equation: upper and lower bounds are given in the
  form of power-laws in 
  time with a gamma exponent corresponding to the first zero-mass symmetric
  harmonic. The bounds bracket the global size of the solution (as measured
  by the supremum) indicating that it decreases in time in the same way as
  that harmonic. However, there does not seem to be any extension of those
  results to prove the actual convergence, for instance, in some L$^p$ metric, of the
  difference between the solution of the PME equation and the first harmonic.

\section{Summary and discussion}\label{sec:discussion}

In the present paper, we first deal with the problem of ambipolar
diffusion of an axial magnetic field, imposing the condition of cylindrical
symmetry around the central axis of the domain, $\Bvec\, =\, B(r)\,
\uvec{z}$. 
By solving an eigenvalue problem, we calculate explicit solutions that are self-similar, have compact support, and pass through a
finite number of nulls. The resulting eigenfunctions have an intrinsic
singularity at the nulls, so phase-plane techniques have been used to
calculate the passage through them. The singularities constitute a finite
collection of sharp current sheets through which magnetic flux is transported
in spite of the vanishing magnetic field thanks to the infinite field
gradient. The existence and various properties of this type of eigenfunction
had been proved in the applied mathematics literature devoted to the porous
medium equation \citep{Hulshof_1991}; to our knowledge, their explicit shapes
have not been published yet.  We can compare the form of the solutions in
the neighbourhood of the singularities calculated in the present paper
(Equation~\ref{eq:current_sheet_solution}) with the properties of the
solutions of the PME: both \citet{Grundy1979} and \citet[][Chapter 16, and
  references therein]{Vazquez_2007} study in detail the nature of the
critical points for the spatial part of the self-similar solutions of the PME
in 1D, 2D, or 3D. These authors consider a phase-plane version of the
equation; \citet{Grundy1979}, in particular, provides a list of all possible
critical points in the problem and a table with a classification of the
solution segments linking those critical points. We find that the individual
lobes of our eigenfunctions can be matched to three different categories in
his table (namely, to those of the third, fourth and fifth rows on page~276,
corresponding, respectively, to (a) the link between our central point and
the first internal null; (b) the link between the outer front and the nearest
internal null; and (c) the link between two internal nulls (for the second or
higher harmonics).  It can be seen that those critical points possess
singularities of the same order as those found in our text.

In the second part of the paper, we propose the set of self-similar
solutions as tests for MHD numerical codes with ambipolar diffusion
capabilities. To show their usefulness and validity, a battery of tests was carried out for the Bifrost code in two spatial dimensions
  starting from initial conditions with cylindrical symmetry
(Sect.~\ref{sec:tests}). We showed that the ambipolar diffusion module
in Bifrost can cope with the passage of the solutions through the current
sheets, with the level of  accuracy increasing the higher the spatial resolution
and in spite of the intrinsic singularity in them. Vice versa, the
tests show that these functions can probe the capabilities of ambipolar
diffusion modules to a larger extent than the simple ZKBP solution that has
been used thus far \citep[e.g.][]{Masson_etal_2012, Vigano2019,
  Nobrega-Siverio_etal_2020b}.  As test functions, the various harmonics
proposed in our paper have the comparative advantage that they combine the $B
\propto |\,\delta\,|^{1/3}$ singularity at the internal nulls (with $\delta$
the distance to the null) with the finiteness of the nonlinear diffusive
flux, and this combination must be sufficiently well reproduced by the code
if it is to pass the test. The ZKBP solution, instead, has a null just at the
outer front, and the singularity there is of a lower order ($\propto
|\delta|^{1/2}$), with zero diffusive flux across it.
On the other hand, the scarcity of tests for the ambipolar diffusion term
until now is in contrast with, for instance, the case of HD shocks, for which a whole
category of exact solutions is available (the solutions of the Riemann
problem) that have been used to develop sophisticated numerical schemes and
tests \citep[see][]{laney_1998, Toro_book}. The contrast to shocks, in fact,
is interesting because of the differences in their mathematical and physical
nature: in shocks, it is the (magneto)hydrodynamic evolution of the
hyperbolic components of the PDE that leads to the formation of the
singularities, which is then smoothed through simple diffusive phenomena
(typically viscosity). In the ambipolar diffusion problem, wherever there is
a null, it is the diffusive phenomenon itself that creates and maintains the
singularity.

The shape of the solutions near the nulls will be modified when the problem
also includes standard Ohmic diffusion (or any other linear diffusion term
for the magnetic field or artificial diffusion designed for the smooth
operation of the code). Let us consider a problem that includes Ohmic
diffusivity in addition to the ambipolar term, but in which the latter is
dominant in a given region. Under such circumstances, the approach to the
singularity is expected to have the standard nonlinear diffusion profile $f
\propto |\,\delta\,|^{1/3}$, except in the immediate neighbourhood of the
null where even a small amount of linear (e.g. Ohmic) diffusivity will
prevent an infinite slope. \citet{Parker63} considered that problem in the
case of a stationary situation \citep[see also][]{Cheung_Cameron_2012}. If
the slope of the field is not infinite at the null, then the diffusive flux
$-\chiamb\, B^2\,B^\prime$ will be zero there. However, the total diffusive
flux (Ohmic plus ambipolar), $-B^\prime/(\muzero\,\sigma) - \chiamb\,
B^2\,B^\prime$, will be nearly uniform around the null and fixed by the
conditions imposed by the dominant agent outside of the null, that is, the
ambipolar diffusion.  For the time-dependent, self-similar solutions of the
present paper, we expect a similar situation: the slope of the function will
be finite in the neighbourhood of the null (as forced by the simple Ohmic
diffusion), but the diffusive flux at the null will be as given by the pure
ambipolar problem. The time evolution of the global solution is therefore
expected to be quite close to that of the pure ambipolar problem.
A behaviour of the numerical solutions around the null as just described also
fits the results of the numerical tests in Sect.~\ref{sec:tests}, even
though the hyperdiffusion terms used by the Bifrost code are not of the
simple Ohmic diffusion kind \citep[see][]{Nobrega-Siverio_etal_2020b}. We
have seen that for a sufficient spatial resolution, the accuracy of the
exponent in the power laws is high and, more to the point, the integral of
the diffusing variable (the magnetic flux) in the central lobe
follows  the analytical law of time evolution
quite precisely (Figure~\ref{fig:power_laws_cylindrical}). That time dependence is physically
and 
mathematically determined by the rate of exchange of magnetic flux across the
nulls, which is given by the diffusive flux. If the presence of
hyperdiffusivities (or Ohmic diffusivities) changed the diffusive flux by
any important amount, the power-law behaviour would not be fulfilled.

The final point in this discussion concerns the asymptotic behavior
  in time (i.e. for large values of $\tau_d$) of the eigenfunctions presented in this
  paper (Sect.~\ref{sec:asymptotics}). By using the Bifrost code, we
  checked that a small enough perturbation (e.g. due to the discretisation error
  of the initial condition and the accumulated error through the update in
  time) does not bring the eigenfunctions out of their initial shape at least
  until a very advanced diffusive time, $\tau_d$, except for a small amount
  that depends on the spatial resolution used for the numerical
  calculation. On the other hand, through experiments in which the initial
  condition, while still of zero net flux and with compact support, deviates
  from the exact eigenfunction shape by a non-small amount, we have seen that
  harmonics higher the first one decay into the latter through merging of the
  internal nulls (or, possibly, via the ejection of the outermost ones through the
  outer front). Instead, if that kind of finite perturbation was given to the
  first harmonic, the solution ends up adopting an exact first-harmonic
  shape. This can be of interest from a twofold perspective: on the one
  hand, from the mathematical point of view, our empirical results (obtained
  numerically) suggest the possibility that the first harmonic is an
  attractor for initial conditions with cylindrical symmetry which have zero
  net magnetic flux. On the other hand, from the point of view of testing MHD
  codes, the higher harmonics pose a double challenge when given as initial
  conditions with no finite perturbation to them: the code must be able to
  maintain the shape of the harmonic until a sufficiently long diffusive
  time is reached.
  
To conclude,  as a general recipe, when testing an
MHD code with ambipolar diffusion capabilities, the following steps are
recommended 
\begin{enumerate}
\item first to test for the ZKBP solution, which has been the standard
  practice in the past;  
\item then to check for the first harmonic and see that it keeps its
  theoretical shape for large diffusive times and following the
  standard power laws with sufficient precision.
\item and, finally, to carry out tests with the higher harmonics to
  check (a) whether small numerical perturbations do not bring them out 
of their initial shape and (b) whether explicit, finite initial perturbations
lead them to transform ino the first harmonic. 
\end{enumerate}

In this paper, we have presented a new family of exact self-similar
solutions to the problem of ambipolar diffusion in a cylindrical geometry and
tested them with the Bifrost code. The extension of these results to problems
with different geometries will be provided in a future paper.

\begin{acknowledgements}
The authors are grateful to an anonymous referee for the careful
  reading of the manuscript and useful suggestions.  FMI thanks
Prof.~J.~Hulshof for his guidance concerning the mathematical literature
on the Porous Medium Equation and for comments to the manuscript;
bibliography suggestions by Prof.~J.L.~V\'azquez are also appreciated.  This
research has been supported by the Spanish Ministry of Science, Innovation
and Universities through projects AYA2014-55078-P and
PGC2018-095832-B-I00. The authors are also grateful to the European Research
Council for support through the Synergy Grant number 810218
(ERC-2018-SyG). DNS acknowledges support by the Research Council of Norway
through its Centres of Excellence scheme, project number 262622, and through
grants of computing time from the Programme for Supercomputing. AWH
gratefully acknowledges the financial support of STFC through the
Consolidated grant, ST/S000402/1, to the University of St Andrews.
\end{acknowledgements}

\bibliographystyle{aa} 

\begin{thebibliography}{70}
\expandafter\ifx\csname natexlab\endcsname\relax\def\natexlab#1{#1}\fi

\bibitem[{Alexiades {et~al.}(1996)Alexiades, Amiez, \& Gremaud}]{Alexiades1996}
Alexiades, V., Amiez, G., \& Gremaud, P.-A. 1996, Communications in Numerical
  Methods in Engineering, 12, 31

\bibitem[{Arber {et~al.}(2007)Arber, Haynes, \& Leake}]{Arber_etal_2007}
Arber, T.~D., Haynes, M., \& Leake, J.~E. 2007, The Astrophysical Journal, 666,
  541

\bibitem[{{Ballester} {et~al.}(2018){Ballester}, {Alexeev}, {Collados},
  {Downes}, {Pfaff}, {Gilbert}, {Khodachenko}, {Khomenko}, {Shaikhislamov},
  {Soler}, {V{\'a}zquez-Semadeni}, \& {Zaqarashvili}}]{Ballester_2018}
{Ballester}, J.~L., {Alexeev}, I., {Collados}, M., {et~al.} 2018, \ssr, 214, 58

\bibitem[{{Barenblatt}(1952)}]{Barenblatt1952}
{Barenblatt}, G. 1952, Prikl. Mat. i Mekh, 16, 67

\bibitem[{{Basu} \& {Ciolek}(2004)}]{Basu_Ciolek_2004}
{Basu}, S. \& {Ciolek}, G.~E. 2004, \apjl, 607, L39

\bibitem[{{Basu} \& {Mouschovias}(1994)}]{Basu_Mouschovias_1994}
{Basu}, S. \& {Mouschovias}, T.~C. 1994, \apj, 432, 720

\bibitem[{{Bernis} {et~al.}(1993){Bernis}, {Hulshof}, \&
  {V\'azquez}}]{Bernis_etal_1993}
{Bernis}, F., {Hulshof}, J., \& {V\'azquez}, J. 1993, J.~reine angew Math.,
  435, 1

\bibitem[{{Braginskii}(1965)}]{braginskii_1965}
{Braginskii}, S.~I. 1965, Reviews of Plasma Physics, 1, 205

\bibitem[{{Brandenburg} \& {Zweibel}(1994)}]{Brandenburg_Zweibel_1994}
{Brandenburg}, A. \& {Zweibel}, E.~G. 1994, Astrophys J Lett, 427, L91

\bibitem[{{Brandenburg} \& {Zweibel}(1995)}]{Brandenburg_Zweibel_1995}
{Brandenburg}, A. \& {Zweibel}, E.~G. 1995, Astrophys J., 448, 734

\bibitem[{{Carlsson} {et~al.}(2016){Carlsson}, {Hansteen}, {Gudiksen},
  {Leenaarts}, \& {De Pontieu}}]{Carlsson_etal_2016}
{Carlsson}, M., {Hansteen}, V.~H., {Gudiksen}, B.~V., {Leenaarts}, J., \& {De
  Pontieu}, B. 2016, \aap, 585, A4

\bibitem[{{Cheung} \& {Cameron}(2012)}]{Cheung_Cameron_2012}
{Cheung}, M. C.~M. \& {Cameron}, R.~H. 2012, \apj, 750, 6

\bibitem[{{Choi} {et~al.}(2009){Choi}, {Kim}, \& {Wiita}}]{Choi_etal_2009}
{Choi}, E., {Kim}, J., \& {Wiita}, P.~J. 2009, \apjs, 181, 413

\bibitem[{Cowling(1957)}]{cowling1957magnetohydrodynamics}
Cowling, T. 1957, Magnetohydrodynamics, Interscience tracts on physics and
  astronomy (Interscience Publishers)

\bibitem[{{Crutcher}(2012)}]{Crutcher_2012}
{Crutcher}, R.~M. 2012, \araa, 50, 29

\bibitem[{{Gonz{\'a}lez-Morales} {et~al.}(2018){Gonz{\'a}lez-Morales},
  {Khomenko}, {Downes}, \& {de Vicente}}]{Gonzalez-Morales_2018}
{Gonz{\'a}lez-Morales}, P.~A., {Khomenko}, E., {Downes}, T.~P., \& {de
  Vicente}, A. 2018, \aap, 615, A67

\bibitem[{{Gonz{\'a}lez-Morales} {et~al.}(2020){Gonz{\'a}lez-Morales},
  {Khomenko}, {Vitas}, \& {Collados}}]{Gonzalez-Morales_2020}
{Gonz{\'a}lez-Morales}, P.~A., {Khomenko}, E., {Vitas}, N., \& {Collados}, M.
  2020, \aap, 642, A220

\bibitem[{{Grassi} {et~al.}(2019){Grassi}, {Padovani}, {Ramsey}, {Galli},
  {Vaytet}, {Ercolano}, \& {Haugb{\o}lle}}]{Grassi:2019}
{Grassi}, T., {Padovani}, M., {Ramsey}, J.~P., {et~al.} 2019, \mnras, 484, 161

\bibitem[{{Gressel} {et~al.}(2015){Gressel}, {Turner}, {Nelson}, \&
  {McNally}}]{Gressel_etal_2015}
{Gressel}, O., {Turner}, N.~J., {Nelson}, R.~P., \& {McNally}, C.~P. 2015,
  \apj, 801, 84

\bibitem[{{Grundy}(1979)}]{Grundy1979}
{Grundy}, R. 1979, Quarterly of Applied Mathematics, 37, 259

\bibitem[{{Gudiksen} {et~al.}(2011){Gudiksen}, {Carlsson}, {Hansteen}, {Hayek},
  {Leenaarts}, \& {Mart{\'{\i}}nez-Sykora}}]{Gudiksen_etal_2011}
{Gudiksen}, B.~V., {Carlsson}, M., {Hansteen}, V.~H., {et~al.} 2011, \aap, 531,
  A154

\bibitem[{{Hansteen} {et~al.}(2007){Hansteen}, {Carlsson}, \&
  {Gudiksen}}]{Hansteen_etal_procs_2007}
{Hansteen}, V.~H., {Carlsson}, M., \& {Gudiksen}, B. 2007, in Astronomical
  Society of the Pacific Conference Series, Vol. 368, The Physics of
  Chromospheric Plasmas, ed. P.~{Heinzel}, I.~{Dorotovi{\v{c}}}, \& R.~J.
  {Rutten}, 107

\bibitem[{{Heitsch} \& {Zweibel}(2003{\natexlab{a}})}]{Heitsch_Zweibel_2003a}
{Heitsch}, F. \& {Zweibel}, E.~G. 2003{\natexlab{a}}, \apj, 583, 229

\bibitem[{{Heitsch} \& {Zweibel}(2003{\natexlab{b}})}]{Heitsch_Zweibel_2003b}
{Heitsch}, F. \& {Zweibel}, E.~G. 2003{\natexlab{b}}, \apj, 590, 291

\bibitem[{{Hulshof}(1991)}]{Hulshof_1991}
{Hulshof}, J. 1991, J. Math. Anal. Appl., 157, 75

\bibitem[{{Hulshof} {et~al.}(2001){Hulshof}, {King}, \&
  {Bowen}}]{Hulshof_etal_2001}
{Hulshof}, J., {King}, J., \& {Bowen}, M. 2001, Advances in Differential
  Equations, 6, 1115

\bibitem[{{Kamin} \& {V\'azquez}(1991)}]{Kamin_and_Vazquez_1991}
{Kamin}, S. \& {V\'azquez}, J. 1991, SIAM J.~Math.~Anal., 22, 34

\bibitem[{{Khomenko} {et~al.}(2021){Khomenko}, {Collados}, {Vitas}, \&
  {Gonz{\'a}lez-Morales}}]{Khomenko_etal_2021}
{Khomenko}, E., {Collados}, M., {Vitas}, N., \& {Gonz{\'a}lez-Morales}, P.~A.
  2021, Philosophical Transactions of the Royal Society of London Series A,
  379, 20200176

\bibitem[{{Khomenko} {et~al.}(2017){Khomenko}, {Vitas}, {Collados}, \& {de
  Vicente}}]{Khomenko_etal_2017}
{Khomenko}, E., {Vitas}, N., {Collados}, M., \& {de Vicente}, A. 2017, \aap,
  604, A66

\bibitem[{{Khomenko} {et~al.}(2018){Khomenko}, {Vitas}, {Collados}, \& {de
  Vicente}}]{Khomenko_etal_2018}
{Khomenko}, E., {Vitas}, N., {Collados}, M., \& {de Vicente}, A. 2018, \aap,
  618, A87

\bibitem[{{Kudoh} \& {Basu}(2008)}]{Kudoh_Basu_2008}
{Kudoh}, T. \& {Basu}, S. 2008, \apjl, 679, L97

\bibitem[{Laney(1998)}]{laney_1998}
Laney, C.~B. 1998, Computational Gasdynamics (Cambridge University Press)

\bibitem[{{Leake} \& {Arber}(2006)}]{leake06}
{Leake}, J.~E. \& {Arber}, T.~D. 2006, Astron.~Astrophys., 450, 805

\bibitem[{{Leake} {et~al.}(2005){Leake}, {Arber}, \&
  {Khodachenko}}]{Leake:2005rt}
{Leake}, J.~E., {Arber}, T.~D., \& {Khodachenko}, M.~L. 2005, \aap, 442, 1091

\bibitem[{{Leake} {et~al.}(2014){Leake}, {DeVore}, {Thayer}, {Burns},
  {Crowley}, {Gilbert}, {Huba}, {Krall}, {Linton}, {Lukin}, \&
  {Wang}}]{Leake_etal_2014}
{Leake}, J.~E., {DeVore}, C.~R., {Thayer}, J.~P., {et~al.} 2014, \ssr, 184, 107

\bibitem[{{Leake} \& {Linton}(2013)}]{leake_linton_2013}
{Leake}, J.~E. \& {Linton}, M.~G. 2013, \apj, 764, 54

\bibitem[{{Leenaarts} {et~al.}(2011){Leenaarts}, {Carlsson}, {Hansteen}, \&
  {Gudiksen}}]{Leenaarts_etal_2011}
{Leenaarts}, J., {Carlsson}, M., {Hansteen}, V., \& {Gudiksen}, B.~V. 2011,
  A\&A, 530, A124

\bibitem[{{Leenaarts} {et~al.}(2007){Leenaarts}, {Carlsson}, {Hansteen}, \&
  {Rutten}}]{Leenaarts_etal_2007}
{Leenaarts}, J., {Carlsson}, M., {Hansteen}, V., \& {Rutten}, R.~J. 2007, \aap,
  473, 625

\bibitem[{{Mac Low} {et~al.}(1995){Mac Low}, {Norman}, {Konigl}, \&
  {Wardle}}]{MacLow:1995}
{Mac Low}, M.-M., {Norman}, M.~L., {Konigl}, A., \& {Wardle}, M. 1995, \apj,
  442, 726

\bibitem[{{Mart{\'\i}nez-Sykora}
  {et~al.}(2017{\natexlab{a}}){Mart{\'\i}nez-Sykora}, {De Pontieu}, {Carlsson},
  {Hansteen}, {N{\'o}brega-Siverio}, \&
  {Gudiksen}}]{Martinez-Sykora_etal_2017a}
{Mart{\'\i}nez-Sykora}, J., {De Pontieu}, B., {Carlsson}, M., {et~al.}
  2017{\natexlab{a}}, \apj, 847, 36

\bibitem[{{Mart{\'\i}nez-Sykora} {et~al.}(2012){Mart{\'\i}nez-Sykora}, {De
  Pontieu}, \& {Hansteen}}]{martinez_sykora_etal_partial_ionization_2012}
{Mart{\'\i}nez-Sykora}, J., {De Pontieu}, B., \& {Hansteen}, V. 2012, \apj,
  753, 161

\bibitem[{{Mart{\'\i}nez-Sykora}
  {et~al.}(2017{\natexlab{b}}){Mart{\'\i}nez-Sykora}, {De Pontieu}, {Hansteen},
  {Rouppe van der Voort}, {Carlsson}, \&
  {Pereira}}]{Martinez-Sykora_etal_2017_science}
{Mart{\'\i}nez-Sykora}, J., {De Pontieu}, B., {Hansteen}, V.~H., {et~al.}
  2017{\natexlab{b}}, Science, 356, 1269

\bibitem[{{Mart{\'\i}nez-Sykora}
  {et~al.}(2020{\natexlab{a}}){Mart{\'\i}nez-Sykora}, {Leenaarts}, {De
  Pontieu}, {N{\'o}brega-Siverio}, {Hansteen}, {Carlsson}, \&
  {Szydlarski}}]{Martinez-Sykora_etal_2020_neq}
{Mart{\'\i}nez-Sykora}, J., {Leenaarts}, J., {De Pontieu}, B., {et~al.}
  2020{\natexlab{a}}, \apj, 889, 95

\bibitem[{{Mart{\'\i}nez-Sykora}
  {et~al.}(2020{\natexlab{b}}){Mart{\'\i}nez-Sykora}, {Szydlarski}, {Hansteen},
  \& {De Pontieu}}]{Martinez-Sykora_etal_2020}
{Mart{\'\i}nez-Sykora}, J., {Szydlarski}, M., {Hansteen}, V.~H., \& {De
  Pontieu}, B. 2020{\natexlab{b}}, \apj, 900, 101

\bibitem[{Masson {et~al.}(2012)Masson, Teyssier, Mulet-Marquis, Hennebelle, \&
  Chabrier}]{Masson_etal_2012}
Masson, J., Teyssier, R., Mulet-Marquis, C., Hennebelle, P., \& Chabrier, G.
  2012, The Astrophysical Journal Supplement Series, 201, 24

\bibitem[{{McKee} \& {Ostriker}(2007)}]{McKee_Ostriker_ARAA_2007}
{McKee}, C.~F. \& {Ostriker}, E.~C. 2007, \araa, 45, 565

\bibitem[{{Mestel} \& {Spitzer}(1956)}]{Mestel_Spitzer_1956}
{Mestel}, L. \& {Spitzer}, L., J. 1956, \mnras, 116, 503

\bibitem[{Mitchner \& Kruger(1973)}]{Mitchner:1973}
Mitchner, M. \& Kruger, C.~H. 1973, Partially Ionized Gases, Wiley series in
  plasma physics (Wiley)

\bibitem[{{Ni} {et~al.}(2021){Ni}, {Chen}, {Peter}, {Tian}, \&
  {Lin}}]{Ni_etal_2021}
{Ni}, L., {Chen}, Y., {Peter}, H., {Tian}, H., \& {Lin}, J. 2021, \aap, 646,
  A88

\bibitem[{{Ni} {et~al.}(2015){Ni}, {Kliem}, {Lin}, \& {Wu}}]{Ni_etal_2015}
{Ni}, L., {Kliem}, B., {Lin}, J., \& {Wu}, N. 2015, \apj, 799, 79

\bibitem[{{Ni} {et~al.}(2016){Ni}, {Lin}, {Roussev}, \&
  {Schmieder}}]{Ni_etal_2016}
{Ni}, L., {Lin}, J., {Roussev}, I.~I., \& {Schmieder}, B. 2016, \apj, 832, 195

\bibitem[{{N{\'o}brega-Siverio}
  {et~al.}(2020{\natexlab{a}}){N{\'o}brega-Siverio}, {Mart{\'\i}nez-Sykora},
  {Moreno-Insertis}, \& {Carlsson}}]{Nobrega-Siverio_etal_2020b}
{N{\'o}brega-Siverio}, D., {Mart{\'\i}nez-Sykora}, J., {Moreno-Insertis}, F.,
  \& {Carlsson}, M. 2020{\natexlab{a}}, \aap, 638, A79

\bibitem[{{N{\'o}brega-Siverio}
  {et~al.}(2020{\natexlab{b}}){N{\'o}brega-Siverio}, {Moreno-Insertis},
  {Mart{\'\i}nez-Sykora}, {Carlsson}, \&
  {Szydlarski}}]{Nobrega-Siverio_etal_2020a}
{N{\'o}brega-Siverio}, D., {Moreno-Insertis}, F., {Mart{\'\i}nez-Sykora}, J.,
  {Carlsson}, M., \& {Szydlarski}, M. 2020{\natexlab{b}}, \aap, 633, A66

\bibitem[{{O'Sullivan} \& {Downes}(2007)}]{osullivan2007}
{O'Sullivan}, S. \& {Downes}, T.~P. 2007, \mnras, 376, 1648

\bibitem[{{Padoan} {et~al.}(2000){Padoan}, {Zweibel}, \&
  {Nordlund}}]{Padoan_etal_2000}
{Padoan}, P., {Zweibel}, E., \& {Nordlund}, {\r{A}}. 2000, \apj, 540, 332

\bibitem[{{Parker}(1963)}]{Parker63}
{Parker}, E.~N. 1963, \apjs, 8, 177

\bibitem[{Pattle(1959)}]{Pattle1959}
Pattle, R.~E. 1959, The Quarterly Journal of Mechanics and Applied Mathematics,
  12, 407

\bibitem[{{Popescu Braileanu} \& {Keppens}(2021)}]{Popescu_Keppens_2021}
{Popescu Braileanu}, B. \& {Keppens}, R. 2021, \aap, 653, A131

\bibitem[{{Shu} {et~al.}(1987){Shu}, {Adams}, \& {Lizano}}]{Shu_etal_1987}
{Shu}, F.~H., {Adams}, F.~C., \& {Lizano}, S. 1987, \araa, 25, 23

\bibitem[{{Tomida} {et~al.}(2015){Tomida}, {Okuzumi}, \&
  {Machida}}]{Tomida_etal_2015}
{Tomida}, K., {Okuzumi}, S., \& {Machida}, M.~N. 2015, \apj, 801, 117

\bibitem[{Toro(2009)}]{Toro_book}
Toro, E.~F. 2009, {Riemann solvers and numerical methods for fluid dynamics: a
  practical introduction; 3rd ed.} (Berlin: Springer)

\bibitem[{V\'azquez(2007)}]{Vazquez_2007}
V\'azquez, J. 2007, The Porous Medium Equation. Mathematical Theory, Oxford
  Mathematical Monographs (Oxford: Clarendon Press)

\bibitem[{{Vernazza} {et~al.}(1981){Vernazza}, {Avrett}, \& {Loeser}}]{val1981}
{Vernazza}, J.~E., {Avrett}, E.~H., \& {Loeser}, R. 1981, \apjs, 45, 635

\bibitem[{{Vigan{\`o}} {et~al.}(2019){Vigan{\`o}}, {Mart{\'\i}nez-G{\'o}mez},
  {Pons}, {Palenzuela}, {Carrasco}, {Mi{\~n}ano}, {Arbona}, {Bona}, \&
  {Mass{\'o}}}]{Vigano2019}
{Vigan{\`o}}, D., {Mart{\'\i}nez-G{\'o}mez}, D., {Pons}, J.~A., {et~al.} 2019,
  Computer Physics Communications, 237, 168

\bibitem[{{Wedemeyer} {et~al.}(2004){Wedemeyer}, {Freytag}, {Steffen},
  {Ludwig}, \& {Holweger}}]{Wedemeyer_etal_2004}
{Wedemeyer}, S., {Freytag}, B., {Steffen}, M., {Ludwig}, H.-G., \& {Holweger},
  H. 2004, \aap, 414, 1121

\bibitem[{{Zel'dovich} \& {Kompaneets}(1950)}]{Zeldovich_Kompaneets_1950}
{Zel'dovich}, Y.~B. \& {Kompaneets}, A. 1950, in Collection of Papers dedicated
  to the 70th birthday of A.F.Ioffe (Moscow: Izd. Akad. Nauk. USSR), 61--71

\bibitem[{{Zel'dovich} \& {Raizer}(1967)}]{Zeldovich_book1967}
{Zel'dovich}, Y.~B. \& {Raizer}, Y.~P. 1967, {Physics of shock waves and
  high-temperature hydrodynamic phenomena} (New York: Academic Press)

\bibitem[{{Zweibel}(1994)}]{Zweibel_1994}
{Zweibel}, E.~G. 1994, in NATO Advanced Science Institutes (ASI) Series C, Vol.
  422, NATO Advanced Science Institutes (ASI) Series C, ed. D.~{Lynden-Bell},
  73

\bibitem[{{Zweibel}(2015)}]{Zweibel_2015}
{Zweibel}, E.~G. 2015, in Astrophysics and Space Science Library, Vol. 407,
  Magnetic Fields in Diffuse Media, ed. A.~{Lazarian}, E.~M. {de Gouveia Dal
  Pino}, \& C.~{Melioli}, 285

\bibitem[{{Zweibel} {et~al.}(2011){Zweibel}, {Lawrence}, {Yoo}, {Ji}, {Yamada},
  \& {Malyshkin}}]{Zweibel_etal_2011}
{Zweibel}, E.~G., {Lawrence}, E., {Yoo}, J., {et~al.} 2011, Physics of Plasmas,
  18, 111211

\end{thebibliography}

\end{document}